\documentclass[twocolumn,pra,showpacs,preprintnumbers,amsmath,amssymb,floatfix,superscriptaddress]{revtex4-1}

 \usepackage[utf8]{inputenc}
 \usepackage[colorlinks,
	linkcolor=blue,
	citecolor=blue,
	urlcolor=blue]{hyperref}
\usepackage[dvips]{graphicx}
\usepackage{dcolumn}
\usepackage{bm}
\usepackage{bbm}
\usepackage{dsfont}
\usepackage{amsmath,amsthm,amssymb,ulem}
\usepackage{graphicx}
\usepackage{epstopdf}
\usepackage{psfrag}
\usepackage{mathrsfs}
\usepackage{color}

\newcommand{\id}{\mathds{1}}

\newcommand{\Hdd}[0]{\mH_{\mbox{\tiny dd}}}
\newcommand{\Hint}[0]{\mH_{\mbox{\tiny int}}}

\newcommand{\Utot}[0]{\mathcal{U}_{\mbox{\tiny tot}}}

\newcommand{\thetaM}[0]{\theta{\mbox{\tiny M}}}

\newcommand{\mR}[0]{\mathcal{R}}

\newcommand{\mH}[0]{\mathcal{H}}
\newcommand{\mM}[0]{\mathcal{M}}
\newcommand{\mU}[0]{\mathcal{U}}
\newcommand{\Urot}[0]{\mathcal{U}_{\mbox{\tiny rot}}}

\newcommand{\bB}[0]{\mathbf B}

\newcommand{\bn}[0]{\mathbf n}

\newcommand{\be}[0]{\mathbf e}

\newcommand{\bv}[0]{\mathbf v}
\newcommand{\ba}[0]{\mathbf a}
\newcommand{\bw}[0]{\mathbf w}
\newcommand{\bmm}[0]{\mathbf m}
\newcommand{\bR}[0]{\mathbf R}
\newcommand{\bI}[0]{\mathbf I}

\newcommand{\bz}[0]{\hat{\mathbf z}}
\newcommand{\br}[0]{\mathbf r}
\newcommand{\bsigma}[0]{\boldsymbol \sigma}

\newcommand{\al}[1]{\begin{align} #1 \end{align}}
\newcommand{\spl}[1]{\begin{align}\begin{split} #1 \end{split} \end{align}}
\newcommand{\vvec}[1]{\begin{pmatrix} #1 \end{pmatrix}}

\newcommand{\ket}[1]{\left\vert{#1}\right\rangle}

\newcommand{\thetatilt}[0]{\theta_{\mbox{\tiny tilt}}}
\newcommand{\tr}[0]{\mbox{Tr}}

\newcommand{\pic}[3]{
\begin{figure}[t!]
\includegraphics[width=#3]{#1} \caption  {#2}
\end{figure}
}

\newcommand{\picwide}[3]{
\begin{figure*}[th!]
\includegraphics[width=#3]{#1} \caption  {#2}
\end{figure*}
}

\newcommand{\refe}[1]{Eq.~(\ref{EQ:#1})}
\newcommand{\refs}[1]{Sec.~\ref{SEC:#1}}
\newcommand{\reff}[1]{Fig.~\ref{FIG:#1}}

\usepackage[margin=1.5cm]{geometry}

\newcommand{\affA}{Department of Chemistry, University of California Berkeley, and Materials Science Division Lawrence Berkeley National Laboratory, Berkeley CA}
\newcommand{\affB}{Engineering Product Development Pillar, Singapore University of Technology and Design, 8 Somapah Road, 487372 Singapore}
\newcommand{\affC}{Research Laboratory of Electronics and Department of Nuclear Science \& Engineering, Massachusetts Institute of Technology, Cambridge, MA}

\renewcommand{\emph}{\textit}

\begin{document}
\title{{Selective decoupling and Hamiltonian engineering in dipolar spin networks}}

\author{A.~Ajoy}\thanks{These authors contributed equally to this work.}\affiliation{\affA}
\email{ashokaj@berkeley.edu}
\author{U.~Bissbort}\thanks{These authors contributed equally to this work.}\affiliation{\affC}\affiliation{\affB} 
\author{D.~Poletti}\affiliation{\affB} 
\author{P.~Cappellaro}\affiliation{\affC}

\begin{abstract}
We present a protocol to selectively decouple, recouple, and engineer effective couplings in mesoscopic dipolar spin networks. In particular, we develop a versatile protocol that relies upon magic angle spinning to perform  Hamiltonian engineering. By using global control fields in conjunction with a local actuator, such as a diamond Nitrogen Vacancy center located in the vicinity of a nuclear spin network, both global and local control over the effective couplings can be achieved. We show that the resulting effective Hamiltonian can be well understood within a simple, intuitive geometric picture, and corroborate its validity by performing exact numerical simulations in few-body systems. Applications of our method are in the emerging fields of two-dimensional room temperature quantum simulators in diamond platforms, as well as in dipolar coupled polar molecule networks.
\end{abstract}
\pacs{03.67.Ac, 76.60.-k, 03.67.Lx}
\maketitle


The concept of Hamiltonian engineering has widespread applications in  quantum information processing~\cite{Blatt12}, quantum metrology~\cite{Giovannetti11}, and in the challenge of constructing suitable platforms for quantum simulation, both analog and digital~\cite{Schirmer07,Verstraete09,Kraus08,Ajoy13b}. In essence, it consists of performing operations on a naturally occurring system to give rise to an effective Hamiltonian, which is of fundamental interest or of direct use for the task at hand~\cite{Feynman82,Lloyd96,Cirac2012}. Numerous quantum control techniques developed over several decades, originating from the field of nuclear magnetic resonance (NMR), have been employed as Hamiltonian engineering tools. For instance, they have been used to effectively break (``decouple") interactions, thereby isolating a quantum system from its environment and greatly increasing coherence times~\cite{Viola98}.

A technique with extensive application in NMR systems to tune dipolar couplings by exploiting its symmetry properties is magic angle spinning (MAS), which traditionally comes in two variants: (i) physical magic angle spinning~\cite{Andrew58,Lowe59,Sakellariou05} where the solid sample is physically rotated at high frequency around an axis tilted by the magic angle $\thetaM=\arctan \sqrt 2\approx 54.7^\circ$ relative to a large static external magnetic field $\bB_0$; 	(ii) MAS that is performed in spin space -- by continuous off-resonant RF irradiation (Lee-Goldberg decoupling~\cite{Lee65,Duer04}) or by a successive application of discrete rotation pulses (\textit{tetrahedral} averaging), each of which effectively rotates each spin by a flip angle of $2\pi/m$~\cite{Pines72,mehring72,Pines92}.

The goal of this work is to extend MAS to nanoscale and mesoscopic systems so as to \textit{engineer} (decouple and recouple) the Hamiltonian of dipolar coupled spin networks. 
The key ingredients of our method are (i)  a bang-bang control~\cite{Boscain06} construction, toggling between two piecewise constant magnetic fields $\bB_0$ and $\bB_1$, that can amplify the effects of even a small transverse magnetic field and achieve a spin rotation around  the magic angle $\thetaM$ for fractions $2\pi/m$; and (ii) a spin actuator~\cite{Khaneja07,Borneman12,Taminiau12} that introduces the transverse magnetic field only locally, thus implementing the rotation construction at the nanoscale. 

\pic{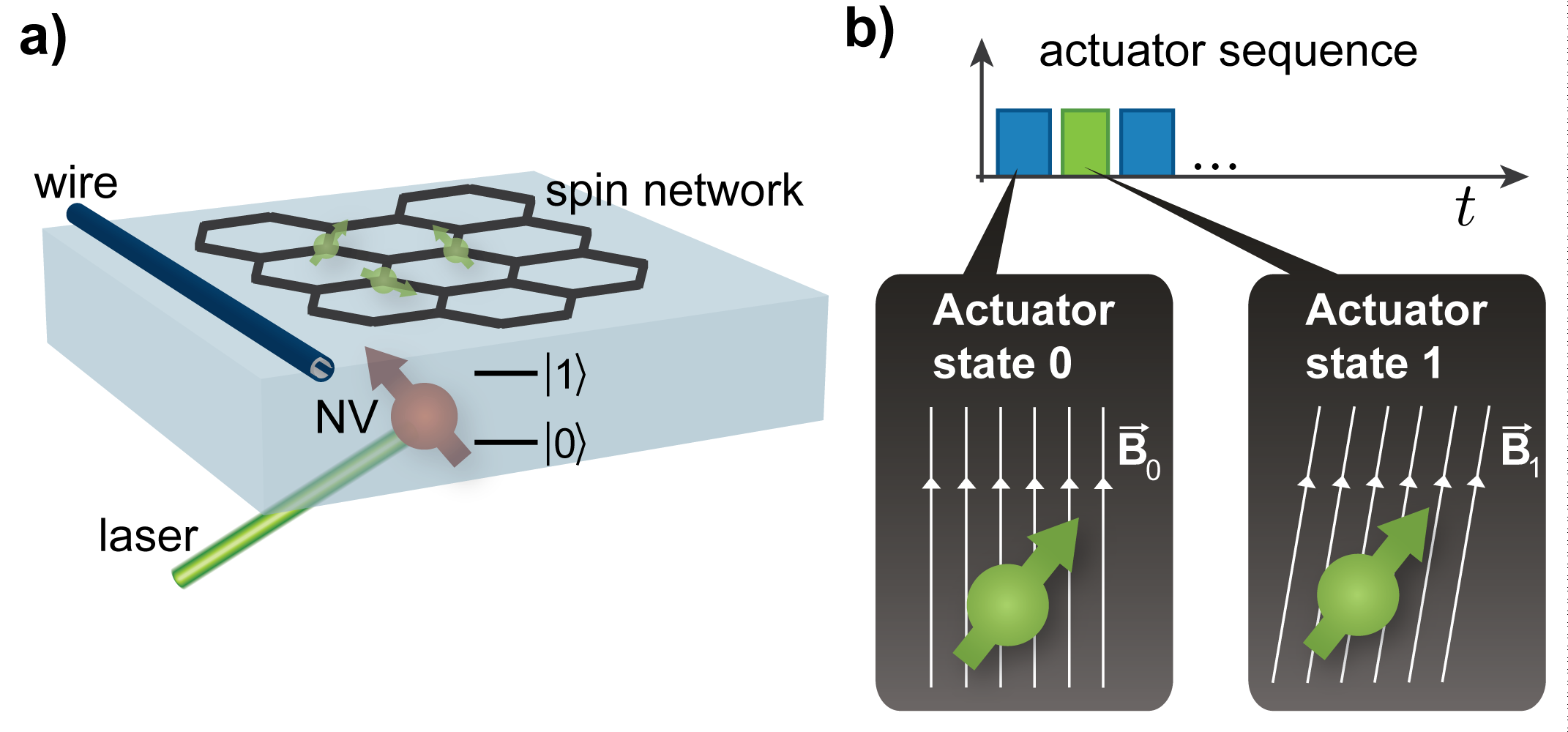}{\label{FIG:Fig1} \textbf{Actuator based selective decoupling and recoupling in spin networks:} (a) A shallow implanted NV center a few nanometers below the diamond surface, optically initialized and controlled by microwave irradiation. (b) By toggling the NV center state, the effective (hyperfine) field for a spin network (shown here for an hexagonal  network eg. Floruographene~\cite{Cai13}, hBN~\cite{Lovchinsky17}) can be turned on and off, realizing two different, discrete Hamiltonians (inset), which can be employed to engineer the effective Hamiltonian of the network (see \reff{Fig4}).}{\linewidth}

\picwide{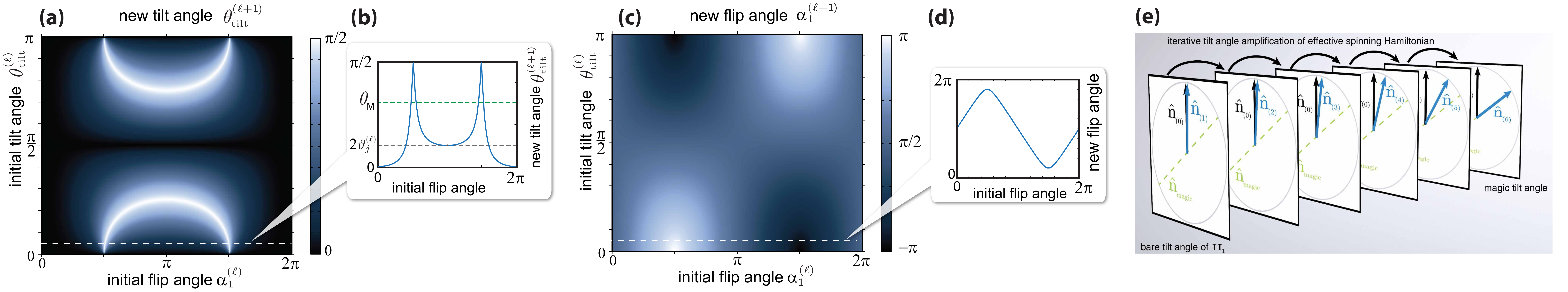}{\label{FIG:tilt_flip_angles} \textbf{Actuator control of spin network:}  Flip angle $\thetatilt^{(l+1)}$ (a) and effective rotation axes flip angle $\alpha_1^{(l+1)}$ (c) after a single iteration of the protocol with slices at a small initial tilt angle $\thetatilt=\pi/10$ are shown in (b) and (d), respectively. In the regimes of high sensitivity [i.e. large slope in (b)] the resulting tilt angle depends strongly on the initial tilt angle and thus the orientation of $\bB_1$, which may lead to spatial selectivity in the vicinity of an actuator. Iterating the protocol, shown in (e), leads to exponential amplification of the effective tilt angle with the blue arrow depicting the effective rotation axis in the respective step.}{\linewidth}

While the technique is more generally applicable to a variety of different experimental realizations, including quantum gases with electric dipolar interactions~\cite{Micheli06,Gorshkov11,Park15}, for clarity we here focus on the specific setup of a shallow Nitrogen Vacancy (NV) center in diamond~\cite{ Morton06,Jelezko06, Borneman12, Suskov13l,Cai13,Ajoy15}  coupled to a nuclear spin network placed above it (\reff{Fig1} a)~\cite{Shi15,Lovchinsky16,Lovchinsky17}. With advances in materials science, these systems have emerged as a promising platform for quantum simulation at room temperature~\cite{Cai13,Wang16,Burgarth17}, and for probing localization and critical phenomena in strongly interacting 2D systems~\cite{Yao14,Nandkishore15,Choi17}. 
Given the small length scales of these quantum networks, they have been traditionally considered hard to control and engineer, even with ultrastrong magnetic gradients~\cite{Degen09,Mamin12}. Instead, as we shall show here, a local actuator up to 10nm away, combined with our new generation of MAS techniques, provides a viable and simple means to engineer these networks at the nanoscale and obtain selective decoupling by four orders of magnitude. The results are underlined by demonstrating the same factor in the growth rate of  entanglement entropy.

In a dipolar network, any two spins $j$ and $j'$ at positions $\br_j$ and $\br_{j'}$ described by spin operators  $\bI^{(j)}=[I_x^{(j)},I_y^{(j)}, I_z^{(j)}]$ and $\bI^{j'}$ interact via the dipolar Hamiltonian $\mH_{j,j'} = D_{j,j'}  (3 (\bI^{j}\cdot \be_{jj'}) (\bI^{j'}\cdot \be_{jj'}) - {\bI^{j}\cdot \bI^{j'}})$, with $D_{j,j'}=-\mu_0  \gamma_j \gamma_{j'} /(8 \pi r_{j,j'}^3)$, where $\gamma_j$ is the gyromagnetic ratio of particle $j$, $r_{j,j'}=|\br_j-\br_{j'}|$ is the distance and $\be_{jj'}=(\br_j-\br_{j'})/r_{j,j'}$ and we set $\hbar=1$. This interaction form applies to both the pairwise interaction between any two spins, as well as to the NV-spin interaction with the NV's gyromagnetic ratio $\gamma_e$ being that of the electron. Since $\gamma_e$ is typically three orders of magnitude larger than that of a nuclear spin $\gamma_n$, the NV-spin interaction at a given distance is larger by the same factor. We envision using the NV center as a local actuator for the spins, by toggling it between two of its internal states $\ket{m=0}$ and $\ket{m=-1}$  either optically~\cite{Gao15} or via microwave pulses~\cite{jelezko04b}. Since the NV's level spacings are much larger than those of the spins, the effect of the spin's states on the NV are negligible and the evolution is well described by only treating the quantum state of the nuclear spins, while assuming the NV to be fixed in one of the two states~\cite{Taminiau12,Kolkowitz12b}. In addition to $\bB_0$ and the local magnetic field created by the NV, a global magnetic field for the spins is toggled on and off by sending piecewise constant pulses of current through a wire located a few $\mu m$ from the spin network~\cite{Ajoy2016b,Lee87}. Hence, spin $j$ is exposed to the magnetic field $\bB_{0/1}^{(j)}$ and evolves under $\mH_{0/1}^{(j)}=\gamma_I \bB_{0/1}^{(j)} \cdot \bI^{(j)}$. Here $\bB_0$ is the global, uniform, static background field, while $\bB_1^{(j)}$  contains, in addition to $\bB_0$, the global contribution from the wire as well as the spin-specific hyperfine (dipolar) contribution from the NV being in the $\ket{m=-1}$ state.

Let us start off with briefly presenting an intuitive description of MAS. Here one assumes idealized, instantaneous rotation pulses $\Urot$. The total evolution over a period of $m$ pulses with the dipolar Hamiltonian $e^{-i \tau \mH_{j,j'} } \Urot e^{-i \tau \mH_{j,j'} } \Urot \ldots$ can then be seen as a sequence of rotated dipolar Hamiltonians $\Urot^m \mH_{j,j'} {\Urot^\dag}^m$ acting sequentially  for $\tau$. Within a Magnus expansion in this frame, and given a dipolar interaction strength much smaller than $\gamma_n \bB_0$, the zeroth order term dominates and leads to an average Hamiltonian $\overline \mH_{j,j'}^{(0)}= \sum_{n=0}^{m-1}  {\Urot}^n   \mH_{j,j'} {\Urot^\dag}^n$. The dipolar interaction in the secular approximation~\footnote{This neglects terms which flip the total $s_z$ in a strong external magnetic field $\bB_0$.} has the particular property that $\overline \mH_{j,j'}^{(0)}$ averages to exactly zero if $\Urot$ rotates around the magic axis, thus effectively decoupling the spins at this order. Moving away from this optimal rotation axis, we expect  residual two-spin interactions in the average Hamiltonian, arising from this zeroth order Magnus term, as shown in \reff{spin_chain}. 

We shall now describe the construction of the effective rotation operator, which allows us to reach the magic angle.
As the elementary building block of our procedure, we let the spin evolve under piecewise constant $\mH_1$, $\mH_0$ and $\mH_1$ for times $\tau_1$, $\tau_0$ and $\tau_1$ respectively, as shown in \reff{Fig1}(b). Our aim, as detailed below, will be to rotate the spins close to the magic angle. 
Omitting the spin label, the evolution of a single spin\footnote{The extension to multiple spins is simply given by taking the direct product of the rotation operators for each spin.} is thus described by the elementary propagator $\mU_{\mbox{\tiny el}} := \mR(\alpha_1, \,  \hat \bn_1 ) \, \mR(\alpha_0, \, \hat \bn_0 ) \, \mR(\alpha_1, \,  \hat \bn_1 )$, where $\mR(\alpha, \, \hat \bn):=e^{-i \alpha {\bI \cdot \hat \bn }}$ is the rotation operator around axis $\hat \bn$ by the flip angle $\alpha$. Being a sequence of rotations, $\mU_{\mbox{\tiny el}}=e^{i\phi_{\mbox{\tiny el}} }\, \mR(\alpha_{\mbox{\tiny el}}, \, \hat \bn_{\mbox{\tiny el}})$ is nothing but a rotation operator itself, up to an unimportant phase $\phi_{\mbox{\tiny el}}$. The functional dependence of the effective axis $\bn_{\mbox{\tiny el}}$ and flip angle $\alpha_{\mbox{\tiny el}}$ on the elementary parameters can be worked out analytically exactly (see App.~\ref{APP:tilt_flip_angle}). Let us summarize the main results: the effective rotation axis  and flip angle are
\al{
\label{EQ:n_eff}
{\mathbf n}_{\mbox{\tiny el}}&=2 b \sin \frac{\alpha_1}{2}  \hat \bn_1 + \sin \frac{\alpha_0}{2} \hat \bn_0\\
\alpha_{\mbox{\tiny el}}&=2\arccos\left(\left|2 b \cos \frac{\alpha_1}{2} - \cos \frac{\alpha_0}{2} \right|\right)
}
respectively, where $b=\cos\frac{\alpha_0}{2} \cos\frac{\alpha_1}{2}-(\hat \bn_0 \cdot \hat \bn_1) \sin\frac{\alpha_0}{2} \sin\frac{\alpha_1}{2}$. Importantly, the rotation axis associated with $\mU_{\mbox{\tiny el}}=e^{-i t \overline \mH}$ and the effective Hamiltonian $\overline \mH$ (see App.~\ref{APP:geometric_pic} for details on the geometric representation of spin Hamiltonians) lie in the plane spanned by the rotation axes $\bn_0$ and $\bn_1$ of the original Hamiltonians, i.e. only the tilt angle is changed and possibly amplified~\cite{Ajoy2017}. In this work, we generally fix $\alpha_0=\pi$. For the case $\alpha_1=\pi$, one finds that the tilt angle is exactly doubled by this protocol. The dependence of the tilt and flip angle of the effective elementary rotation on $\alpha_1$ (see \reff{tilt_flip_angles}) shows distinct regimes of \textit{high} and \textit{low sensitivity} on the initial tilt angle [corresponding to steep and shallow slopes in \reff{tilt_flip_angles}(b)]. 
Perhaps non-intuitively, for any non-zero initial tilt angle, the effective tilt angle can actually reach the magic angle with only one application of $\mU_{\mbox{\tiny el}}$, but at the cost of a very small flip angle [close to $0$ or $2\pi$ in \reff{tilt_flip_angles}(a) and (b)]. Alternatively, by concatenating the procedure $\mU^{(k)} \mapsto \mU^{(k+1)}= \mU^{(k)} \, \mR(\alpha_0, \, \hat \bn_0 ) \, \mU^{(k)}$ with the initial $\mU^{(1)}=\mR(\alpha_1, \,  \hat \bn_1 )$, one can  amplify the tilt angle, exponentially in the number of concatenations (e.g. doubling $\thetatilt^{(l)}$ at $\alpha_1=\pi$ with every in iteration) and linearly in time. Given a small initial bare tilt angle of $\bB_1$ (we use $\thetatilt^{(0)}=2.5^\circ$ throughout this work, well within reach for typical nuclear spin systems~\cite{Cai13,Ajoy15}), a general iteration of the elementary concatenation $\mU^{(N_c)}$ (e.g. for $\alpha_1=\pi$) does generally not exactly lead to the magic angle unless $2^{N_c} \thetatilt^{(0)} =\thetaM$. We therefore introduce a control sequence that gives our final protocol, reaching the desired total unitary
\spl{
\label{EQ:Utot}
\mU_{\mbox{\tiny tot}}=\mU^{(N_c)} e^{-i \tau_b \mH_0} \mU^{(N_c)} e^{-i \tau_a \mH_0} \mU^{(N_c)} e^{-i \tau_b \mH_0} \mU^{(N_c)}.
}

Given that the rotation axis of $\mU^{(N_c)}$ is sufficiently large $\thetatilt^{(N_c)} \in [\thetaM/2, \thetaM]$, the topological structure of the resulting flip and tilt angle of $\mU_{\mbox{\tiny tot}}$ guarantees that suitable values of $\tau_a$ and $\tau_b$ always exist [App.~(\ref{APP:tilt_flip_angle})] to reach exactly the magic angle and a flip angle of $2\pi/m$. However, the point $(\tau_a, \tau_b)$, which optimizes the sequence, is generally not unique (beyond the approximate $2\pi/|\bB_0|$ periodicity) and different points feature different decoupling characteristics and \textit{sensitivity} (i.e. the strength of the effective tilt angle dependence on $\alpha_1$), a feature that allows for further tuning of the effective interactions. The total time required by $\mU_{\mbox{\tiny tot}}$ is $\tau_{\mbox{\tiny rot}}=\tau_{a}+2\tau_{b}+4[2^{N_c} \tau_1 +(2^{N_c}-1) \tau_0]$, which, in contrast to the usual idealized discrete rotation pulses used in MAS~\cite{Pines72}, may be non-negligible on the time scale of the inverse dipolar interactions. To perform the decoupling, we let the system to evolve freely under $\mH_0$ for a wait time $\tau$ between the rotations $\mU_{\mbox{\tiny tot}}$.

\pic{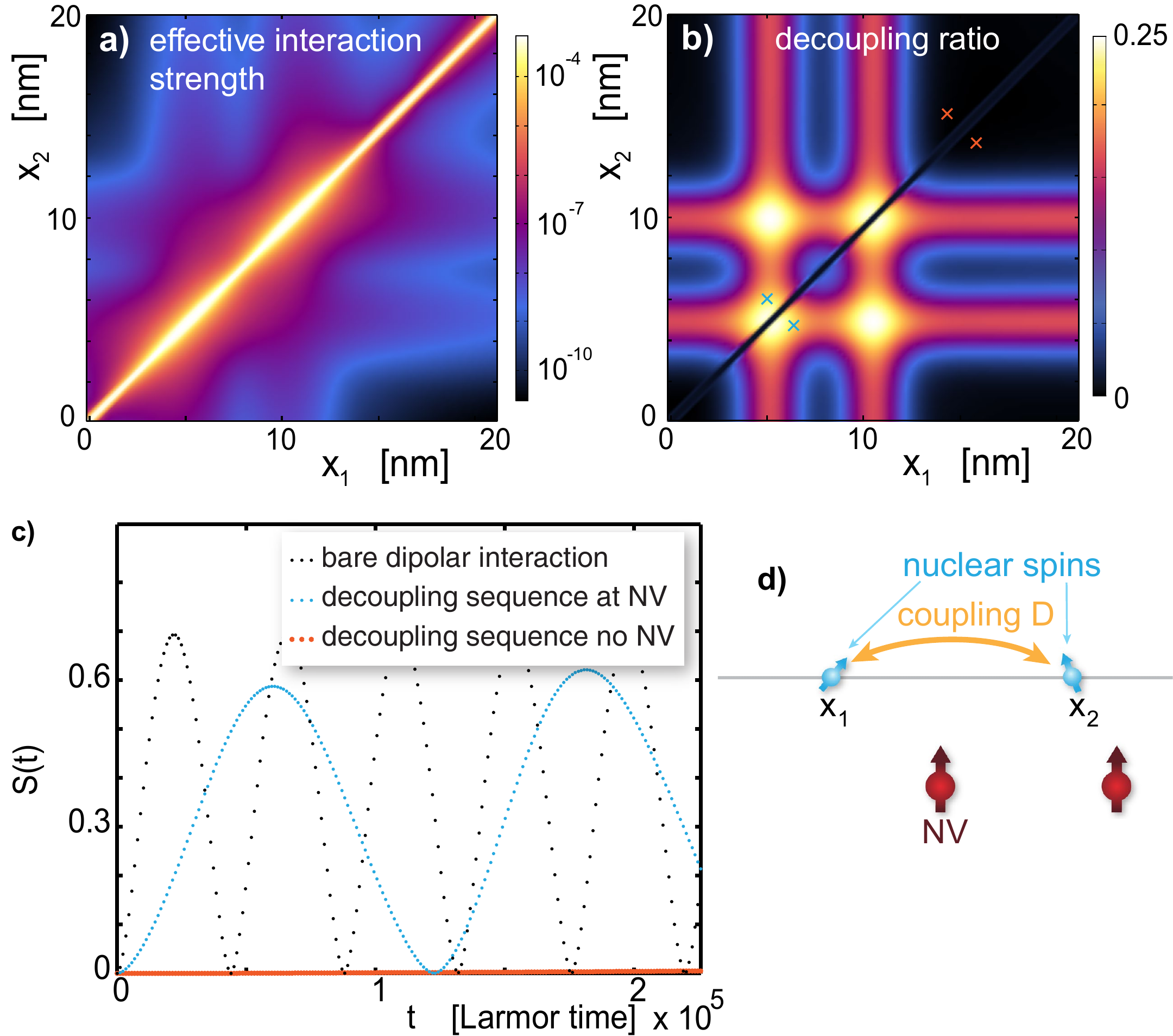}{\label{FIG:spin_chain} \textbf{Decoupling and selective recoupling via local actuators:} By locally disturbing the optimal decoupling field strength using NV centers (as sketched in (d)), the electronic state of which can be toggled in synchrony with the $\bB_1$ magnetic field, the effective interaction between two nuclear spins can be made spatially dependent. This is reflected by both the effective interaction strength $\overline {\mathcal S}_{1,2}$ shown in (a) and the decoupling ratio ${\mathcal R}_{1,2}$ in (b). The extent to which the decoupling and recoupling works is reflected by the temporal growth (shown here stroboscopically) of the entanglement entropy of two spins initially in the $\ket{\uparrow}\otimes\ket{\downarrow}$ state (corresponding parameter regimes indicated by crosses in (b)). Parameters: $\tau_a=0.95118$, $\tau_b=0.93851$, $\tau=1000.5$, $N_c=5$, $m=3$, NV positions $[5,0,-2.5]$nm and $[11,0,-2.5]$nm.}{\linewidth}

\paragraph{Effective Hamiltonian: }
For the effective interaction to be canceled efficiently, one is required to consider $m$ repetitions of rotation by $2\pi/m$. The Hamiltonian is then periodic with period $\tau_{\mbox{\tiny rot}}+\tau$ and the evolution of any state can be described within average Hamiltonian theory~\cite{Haeberlen71} when evaluated stroboscopically. To quantitatively evaluate the decoupling effects on the spin dipolar couplings, we consider the effective (Floquet) Hamiltonian over a period $T=m(\tau_{\mbox{\tiny rot}}+\tau)$ in the toggling frame~\cite{Haeberlen71}, $\overline \mH_I= \frac i T \ln [\mU_I(t_0, t_0+T)]$ [also see App.~(\ref{APP:strength_quantification}) for a detailed discussion]. $\overline \mH_I$ can be decomposed into a series of terms: scalar, single-spin and two-spin interactions. 
In this work, we aim at engineering or canceling a two-spin interaction Hamiltonian. We thus focus on the two-spin terms,  which can be extracted by projecting the operator amplitudes $h_{k,l}^{(j,j')}=\mbox{Tr} \left[ \overline \mH_I ( \sigma_k^{(j)} \otimes \sigma_l^{(j')}  \otimes \prod_{k\neq j, j'} \mathbbm 1^{(k)} ) \right]/4$ on the subset of two-spin operators. To quantify the effective coupling strength, we use the Frobenius norm $\overline {\mathcal S}_{j,j'}=\| \overline \mH_{j,j'} \|_F$ of the associated operator $\overline \mH_{j,j'}=\sum_{k,l} h_{k,l}^{(j,j')} \sigma_k  \otimes \sigma_l $ on the two-spin space (see App.~\ref{APP:strength_quantification}), which is equivalent to the Euclidean norm of the coefficient vector $h_{k,l}^{(j,j')}$. The bare interaction strength ${\mathcal S}_{j,j'}$ can be quantified analogously and for a simple measure of the decoupling procedure's performance, we define the decoupling ratio ${\mathcal R}_{j,j'}=\overline {\mathcal S}_{j,j'}/ {\mathcal S}_{j,j'}$.

\paragraph{Hamiltonian Engineering: }
																																	
To demonstrate our protocol, we first consider decoupling and local recoupling of a nuclear $^{13}$C spin pair, located at positions $\bR_{1/2}=[x_{1/2},0,0]^t$ (\reff{spin_chain}(d)). 
A first remarkable result is that our method can reach high decoupling efficacy. Starting in a low sensitivity regime  given by $\tau_1=\pi/|\bB_0|$,   $N_c=5$ and $m=3$, we numerically optimize the decoupling with respect to $(\tau_a, \tau_b, \tau)$ (in the absence of the NVs) by initializing the minimization close to the half integer decoupling peak at $\tau=1000.5 \times 2 \pi/(\gamma_{\mbox{\tiny I}} |\bB_0|)$. Using these optimized parameters, we show the resulting interaction strength $\| \overline \mH_{j,j'} \|_F$ and decoupling ratio ${\mathcal R}_{j,j'}$ in \reff{spin_chain}(a) and (b) respectively. Interestingly, if the spins are apart farther than $\approx 0.2\mbox{nm}$,  ${\mathcal R}_{j,j'}$ converges to a constant, i.e. becomes distance independent and isotropic. This also holds for parameters away from high decoupling. 
Decoupling by three to four orders of magnitude are reached in \reff{spin_chain} (nine orders of magnitude or more can be reached when optimizing around other decoupling peaks), a non-trivial result given the highly non-ideal nature of the pulses compared to the typical assumptions of instantaneous pulses (see \refs{wait_time} for a comparison) entering the intuitive explanation of MAS. The high efficacy of the decoupling is corroborated by the strongly suppressed growth of the entanglement entropy of two spins initially prepared in the uncorrelated state $\ket{\uparrow}\otimes\ket{\downarrow}$, shown in \reff{spin_chain}(c) as orange dots, as compared to the same system without the protocol (black dots).

\picwide{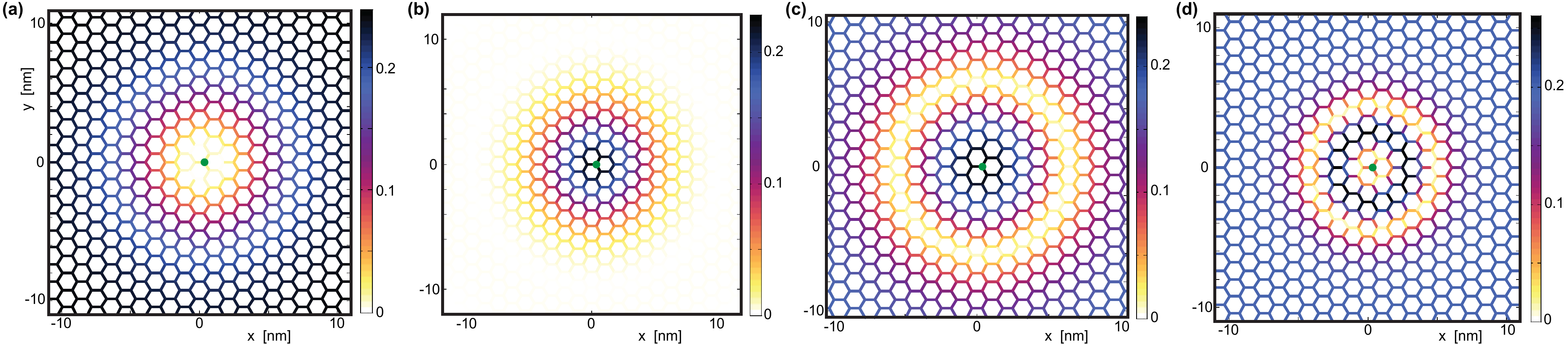}{\label{FIG:Fig4}\textbf{Selective decoupling in hexagonal spin networks:} Qualitatively different spatial coupling topologies can be created within our protocol in conjunction with a local actuator (e.g. an NV center) located several nanometers (between $5.7$nm and $10$nm here) below a $^{13}$C hexagonal spin lattice. 
The decoupling ratio $\mathcal R_{j,j'}$ (directly proportional to the effective interaction strength $||\overline \mH_{j,j'} ||$, as the nearest neighbor distance is constant) for all nearest neighbor spin pairs is shown in color-coded form with the spins located at the respective nodes. Depending on the chosen parameters, the decoupling protocol has a different length scale which can be exploited to give rise to different coupling structures. For instance, one can create a \textit{non-interacting patch} (\textbf{a}) within an otherwise interacting network or vice versa, create a small \textit{interacting cluster} in an otherwise essentially non-interacting lattice as shown in (\textbf{b}) by tuning the protocol to decouple the spins in the absence of the local actuator. Moreover, the parameters can be tuned to separate off an \textit{interacting patch} shown in (\textbf{c}), or create an effective \textit{1D ring-like lattice topology} as shown in (\textbf{d}). For all subfigures $N_c=1$, the NV-center is located at $(x=0.355\mbox{nm}, y=0)$ (marked by a green circle) with variable depth $z$ and the $^{13}$C spins are located on a regular hexagonal sublattice with lattice constant $a_{\mbox{\tiny lat}}=5 a_{\mbox{\tiny cc}}$, where $a_{\mbox{\tiny cc}}=0.142nm$ is the carbon-carbon bond length and the lattice constant of graphene. We choose this sublattice configuration to obtain a system in a \textit{weakly interacting regime}, i.e. allowing the true decoupling ratio to be obtained from a pairwise treatment of spins and avoid many-body effects. A relative agreement of $0.5\%$ or less is found with a six spins cluster calculation (including the nearest neighboring spins). Furthermore, for all of these plots the NV state is activated throughout the entire wait period $\tau$ which greatly amplifies the effect of the NV on $\overline \mH$. For a complete list of parameters used, see App.~(\ref{SEC:parameters_hexagonal_plots}).
}{\linewidth}

We now discuss the central result of our work, recoupling via the NVs acting as local actuators. As reflected by ${\mathcal R}_{j,j'}$ in \reff{spin_chain} (b) and the entanglement entropy growth in \reff{spin_chain} (c), the NV, acting as a local actuator several nanometers from the nuclear spins, can disrupt the protocol in the sensitive regime and recouple the spin interaction to almost the original strength. We emphasize that the strong recoupling occurs despite the NVs' magnetic field contribution being three orders of magnitude smaller than $\bB_0$. This is again corroborated by the rapid growth of the entanglement entropy (blue dots) in \reff{spin_chain} (c) relative to the decoupled case (orange dots).  The localized nature of the NVs breaks the translational symmetry of the inter-spin interaction, making the latter a function of the coordinates of both spin and yielding a different scaling from the typical  $|\br -\br'|^{-3}$ dipolar interaction form. Furthermore, the structure of $\overline \mH$ is fundamentally different: whereas $\mH_{j,j'}$ is always of an $XXZ$-form in a suitably rotated basis, $\overline \mH_{j,j'}$ can be of a more general form. We remark that if the rotation axis of $\mU_{\mbox{\tiny el}}$ lies on the magic axis, we always find decoupling (App.~\ref{APP:decoupling_peaks}), i.e. $\mathcal{S}_{j,j'}$ essentially vanishes. This underlines that our procedure can indeed be explained with MAS.

A further prediction made by the Magnus expansion can be corroborated to a high degree in exact simulations: The independence of $\mR_{j,j'}$ on the position of the spins under global pulses if the spins are sufficiently separated such that $|\mH_{j,j'}|\ll |\gamma_n \bB_0|$, which holds up to a small inter-spin distance, e.g. $\approx 0.12$nm for the $^{13}$C spins.

\paragraph{Selective Decoupling in 2D Hexagonal Lattice: }
Next, we demonstrate selective decoupling on a hexagonal lattice of nuclear spins, as realized in graphene or 2D HBN above a shallow NV in  diamond in a recent single spin NMR experiment \cite{Lovchinsky16}. 
In this system, we demonstrate the selective local decoupling of a spin cluster, as well as the creation of a locally coupled subnetwork, which is decoupled from the remaining network. Both constitute important tools in the context of realizing locally addressable quantum memories.

To achieve this, we pick a spin pair $j,j'$, for which we optimize the decoupling procedure. First, we pick an approximate $\tau_1$, which sets the sensitivity (see \reff{tilt_flip_angles}) and the number of required concatenations of the protocol. Thereafter, we pick a decoupling peak in the $\tau_a, \tau_b$ space. Using these initial values ($\tau_a, \tau_b, \tau$) close to the true minimum, we optimize $\mR_{j,j'}$ in the full spin network as a function of these parameters. This sets the timings for the full network and thus fully determines $\overline \mH$. Within this numerically exact network simulation, the decoupling works surprisingly well, often leading to decoupling ratios $\mR<10^{-10}$ for the optimized pair. 
In \reff{Fig4} the decoupling ratio (proportional to the effective interaction strength) between all nearest neighbor pairs are shown for spins arranged on a sublattice of a graphene sheet. By carefully choosing the timings and NV position relative to the decoupling pair $j,j'$ and the lattice symmetry points, various scenarios are possible: a local \textit{non-interacting patch} can be \textit{burned} into the lattice (\reff{Fig4} (a)). This may be useful for selectively preserving the qubits' states and act as a temporary quantum memory. Other decoupling topologies are also possible by choosing different parameters: for instance in \reff{Fig4} (b) and (c) small interacting patches or ring-like topologies (\reff{Fig4} (d)), decoupled from the rest of the lattice, can be created. These result are obtained for $^{13}$C nuclear spins spaced at a large distance on a graphene sublattice, where the system is weakly interacting. By choosing a smaller inter-spin distance, the system can be brought into a strongly interacting regime, where the effective interaction between a given pair of spins also strongly depends on the neighboring spins, leading to a highly non-trivial system beyond the scope of this work.

We conclude with a few short comments. In the protocol presented above, one can easily switch between the two different effective Hamiltonian regimes shown in \reff{Fig4} (a) and (b)  during a single experimental run by simply changing the protocol timings. Effective many-particle interactions (e.g. three spin terms) also arise in $\overline \mH$, and these are also well captured and explained in higher orders of the Magnus expansion. However, they are much smaller (typically $10^{-2}$ to $10^{-3}$) than the two spin terms, but may become important for the dynamics if the latter vanish. Furthermore, adding a tertiary spin to a given pair, not only do three-spin terms arise, but the presence of the tertiary spin can also induce an effective two-spin interaction $\overline \mH_{j,j'}$ not present in the bare two-spin system (see \refs{tertiary_spin_induced_int}).

In summary, we have described a new versatile method for selective Hamiltonian engineering of dipolar spin networks using local actuator control. Our method relied on a modified version of magic angle spinning, but carried out by global fields in conjunction with spin actuators that provide a sensitive way of decoupling or recoupling interactions with tunable range of action. While more generally applicable, we demonstrated the protocol to the specific case of nuclear spin networks placed over the surface of diamond containing shallow implanted NV centers that act as the actuators. This technique presents a compelling strategy for nanoscale Hamiltonian engineering and opens the path for quantum simulators constructed out of dipolar networks.

{\it Acknowledgments}:
We gratefully thank A. Pines for insightful conversations. U.B. and D.P. acknowledge support by the Air Force Office of Scientific Research under Award No.FA2386-16-1-4041. P.C. acknowledges support from NSF under PHY1734011 and PHY1415345, and ARO under W911NF-15-1-0548.

\bibliography{main}
\bibliographystyle{apsrev4-1}

\clearpage

\begin{appendix}

\section{Construction for MAS under bang-bang control}
\label{APP:tilt_flip_angle}
\textit{Effective tilt and flip angles under the pulse sequence} --

Given a sequence of rotations $e^{i\phi_{\mbox{\tiny tot}} }\mR(\alpha_1, \,  \bn_1 ) \, \mR(\alpha_0, \, \bn_0 ) \, \mR(\alpha_1, \,  \bn_1 )$ we seek the collective flip and tilt angles. These can be expressed analytically by, first, writing each partial rotation in the form
\al{
\label{EQ:rot_op_rel}
e^{-i \alpha {\vec \sigma \cdot \bn }/{2}}=\cos{\frac \alpha 2} \id - i ( \bn \cdot \vec \sigma ) \sin {\frac \alpha 2}.
} 
Simplifying them using the relations $( \bn_0 \cdot \vec \sigma)(\bn_1 \cdot \vec \sigma)=\bn_0 \cdot \bn_1+ i \vec \sigma \cdot(\bn_0 \times \bn_1)$ and $(\bn_1 \times \bn_0) \times \bn_1 = \bn_0 - \bn_1 (\bn_1 \cdot \bn_0)$ results in, 
\al{
\mU_{\mbox{\tiny tot}}=& \left(2 b \cos \frac{\alpha_1}{2} - \cos \frac{\alpha_0}{2} \right)\id \\
& - i\left[ 2 b \sin \frac{\alpha_1}{2}  \bn_1 + \sin \frac{\alpha_0}{2} \bn_0 \right] \cdot \vec \sigma
}
with
\al{
b= \cos \frac{\alpha_0}{2}\cos \frac{\alpha_1}{2} - (\bn_0\cdot \bn_1) \sin \frac{\alpha_0}{2} \sin \frac{\alpha_1}{2}.
}
By comparison with Eq.~(\ref{EQ:rot_op_rel}), the total effective flip angle is 
\al{
\alpha_{\mbox{\tiny tot}}=2\arccos\left(\left|2 b \cos \frac{\alpha_1}{2} - \cos \frac{\alpha_0}{2} \right|\right)
}
and the effective rotation axis is $\hat \bn_{\mbox{\tiny tot}}={\mathbf n}_{\mbox{\tiny tot}}/| {\mathbf n}_{\mbox{\tiny tot}}|$ with 
\al{
{\mathbf n}_{\mbox{\tiny tot}}&=2 b \sin \frac{\alpha_1}{2}  \hat \bn_1 + \sin \frac{\alpha_0}{2} \hat \bn_0.
}

\textit{Final Concatenation Step to Reach the Magic Angle} --

\begin{figure*}[hbt!]
  \centering
  {\includegraphics[width=\linewidth]{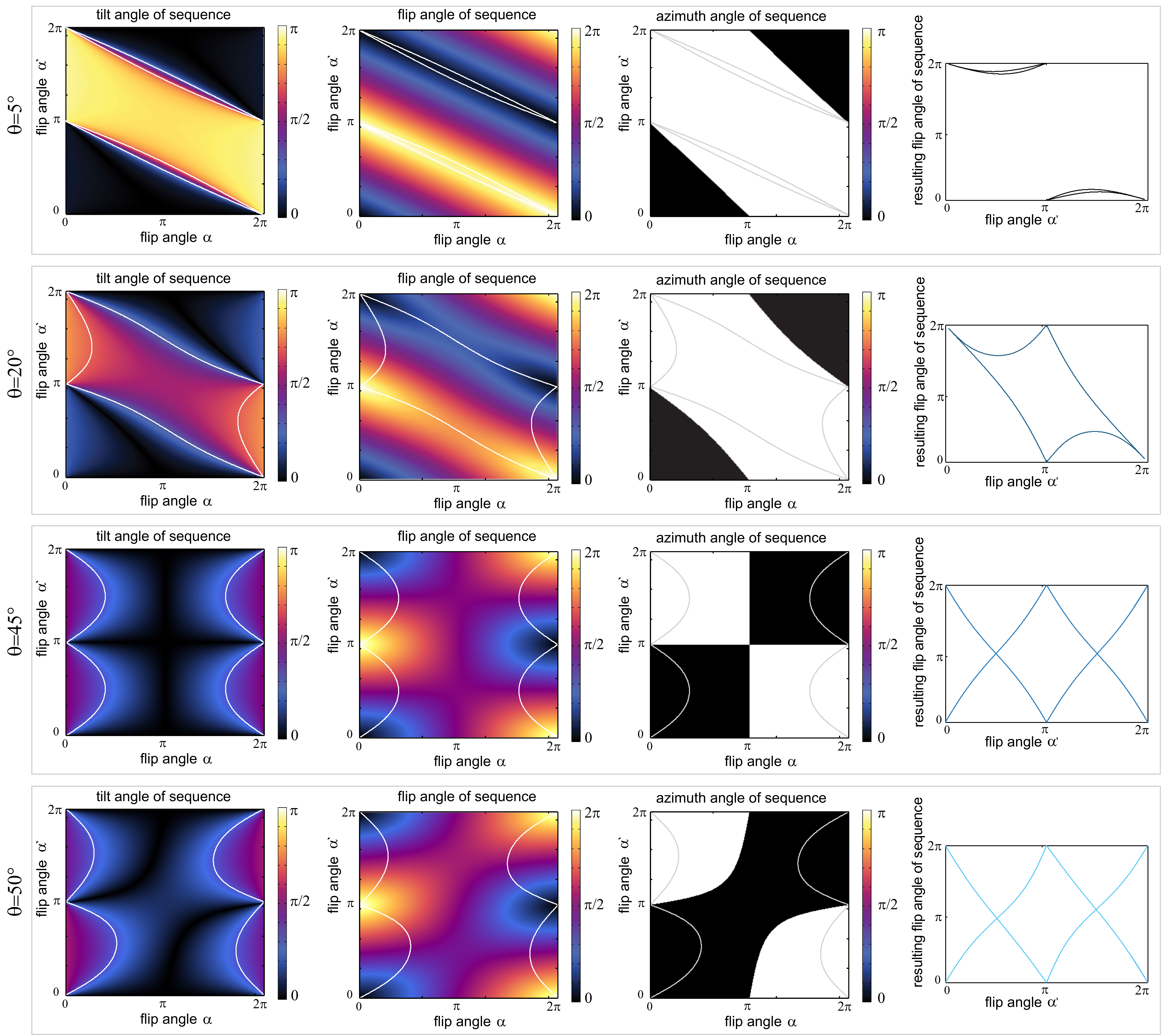}}
  \caption{ \label{FIG:combined_final_pulse_seq}
The tilt angle (left column), flip angle (second column) and azimuth angle (third column) for the entire pulse sequence in \refe{Utot} for various bare tilt angles $\theta_{\mbox{\tiny bare}}$ shown in the different rows. The white lines indicate the regions where the effective Hamiltonian is tilted at the magic angle, i.e. $\theta_{\mbox{\tiny tot}}=\thetaM$ or $\theta_{\mbox{\tiny tot}}=\pi-\thetaM$. We point out the symmetry that as $\theta_{\mbox{\tiny bare}} \mapsto \frac \pi 2 - \theta_{\mbox{\tiny bare}}$, each of these functions is mirrored upon the vertical line $\alpha=\pi$. Therefore, the $\theta_{\mbox{\tiny bare}}=45^\circ$ row has a reflectional symmetry throughout. The large values of tilt angles $\alpha$ and $\alpha'$ can be generated by the concatenation sequence in \refe{Utot}. The fourth column shows the flip angles at the magic angle as a function of $\alpha'$. This is obtained by, for given $\alpha'$, searching for the value of $\alpha$ at which $\theta_{\mbox{\tiny tot}}=\thetaM$ and determining the flip angle at that point. For each $\alpha'$ two such values exist.
	}
\end{figure*}

Given an initial tilt angle $\theta_{\mbox{\tiny tilt}}^{(0)}=\arccos(\hat \bn_0 \cdot \be_z)$ of the Hamiltonian $\mH_1$, the iterative concatenation $\mU^{(k+1)}=\mU^{(k)} \mR(\alpha_0, \hat \bn_0) \mU^{(k)}$ essentially doubles the tilt angle with every step, while keeping the flip and azimuth angles fixed at $\pi$ and $0$ respectively. However, one can only reach the magic angle $\thetaM$ with this procedure if a $k\in\mathbb N$ exists, such that $\theta_{\mbox{\tiny bare}} \, 2^k =\theta_M$. As this is generally not the case, we here discuss a modification of the procedure for the final step to still reach the magic angle exactly with vanishing azimuth angle $\varphi_{\mbox{\tiny tot}}$ of the effective rotation axis, at the cost of a possibly modified flip angle $\alpha_{\mbox{\tiny tot}}\neq \pi$. We start with the sequence 
\spl{
\label{EQ:Utot_general}
\mU_{\mbox{\tiny tot}}=\mU^{(N_c)} e^{-i \tau_b \mH_0} \mU^{(N_c)} e^{-i \tau_a \mH_0} \mU^{(N_c)} e^{-i \tau_c \mH_0} \mU^{(N_c)}.
}
Being a rotation, $\mU_{\mbox{\tiny tot}}$ can again be parametrized using three real numbers: the flip angle $\alpha_{\mbox{\tiny tot}}$, and effective rotation axis $\hat \bn_{\mbox{\tiny tot}}$ parametrized by an effective tilt angle $\theta_{\mbox{\tiny tot}}$ and an azimuth angle $\varphi_{\mbox{\tiny tot}}$, which are functions of of $(\tau_a,\tau_b, \tau_c)$. As before, we wish to keep the axis $\hat \bn_{\mbox{\tiny tot}}$ in the same plane as the axis of $\mU^{(N_c)}$, i.e. $\varphi_{\mbox{\tiny tot}}=0$ or $\varphi_{\mbox{\tiny tot}}=\pi$, which can be achieved by setting $\tau_c=\tau_b$ (as reflected in the third column of \reff{combined_final_pulse_seq}), thus eliminating one variable and leads from \refe{Utot_general} to \refe{Utot} of the main paper.

We now analyze the dependence of the resulting $\mU_{\mbox{\tiny tot}}$ restricted to this plane on the angles 
\al{
\alpha&=\gamma_I |\bB_0| \tau_a\\
\alpha'&=\gamma_I |\bB_0| \tau_b
}
and will show that any set of values
 $(\alpha_{\mbox{\tiny tot}}, \, \theta{\mbox{\tiny tot}} )$ can be reached once the bare tilt angle $\theta=\theta_{\mbox{\tiny bare}} \, 2^k$ lies above a critical threshold. Importantly, the tilt angle $\theta{\mbox{\tiny tot}}$ can reach the magic angle with an arbitrary flip angle $\alpha{\mbox{\tiny tot}}$.

In the left column of \reff{combined_final_pulse_seq} we show the resulting tilt angle as a function of the remaining variables $(\alpha,\, \alpha')$. At the points $(\alpha, \, \alpha')$ with $\alpha \in \{0, \, 2\pi \}$ and $\alpha \in \{0, \, \pi, \, 2\pi \}$, both $\theta{\mbox{\tiny tot}}$ and $\alpha{\mbox{\tiny tot}}$ feature some interesting behavior: as a function of $\alpha'$, the total tilt angle develops a discrete jump between the values $0$ and $2\pi$ in the limit of $\theta \to 0$ (the limit of going upwards in the rows of \reff{combined_final_pulse_seq}). Equipotential lines of different value enter these points at different angles in the plane, such as the one highlighted in white corresponding to the values of the total tilt angle taking on the magic angle $\thetaM$.  This line evolves continuously as the external $\theta$ (different rows) is varied, but undergoes two transitions where the topology of the curves changes qualitatively (between the first and the second, as well as the second and the third row). 

The second column shows the flip angle in the same $(\alpha, \, \alpha')$-plane and we also draw the manifold where $\theta{\mbox{\tiny tot}} =\thetaM$ in white. We note that at the points $\alpha \in \{0, \, 2\pi \}$ and $\alpha \in \{0, \, \pi, \, 2\pi \}$ this function $\alpha{\mbox{\tiny tot}}$ takes on values of zero and $2\pi$ in alternating order and varies continuously throughout the entire plane. Thus, in the second and third regime (second to fourth row in \reff{combined_final_pulse_seq}), the magic tilt angle manifold traverses all possible values of the flip angle, as shown more explicitly in the fourth column of \reff{combined_final_pulse_seq}. In these regimes, with the corner points of the manifold fixed to zero and $2\pi$, there is thus a line connecting the upper and lower points. Hence,  the reachability of an arbitrary flip angle is topologically guaranteed once a sufficiently large initial $\theta$ is given.

We point out that the sensitivity of the flip angle $\alpha_{\mbox{\tiny tot}}$ remains low through the entire region and for arbitrary value of $\theta$, whereas the strong selectivity of the tilt angle $\theta_{\mbox{\tiny tot}}$ can be seen for small $\theta_{\mbox{\tiny tot}}$ (upper left subfigure of \reff{combined_final_pulse_seq}).

\section{Geometric Hamiltonian Representation}
\label{APP:geometric_pic}
To obtain a more intuitive understanding of the procedure, we now introduce a geometric representation for the (effective) Hamiltonians. Any given single spin Hamiltonian \footnote{Up to an unimportant total energy shift proportional to the unit operator.} can be expressed in terms of a 3D vector $\bn_H$ as $H=\bn_H \cdot \bsigma$, where $\bsigma=[\sigma_x,\,\sigma_y,\,\sigma_z]^t$ is the vector of Pauli operators and for a Hermitean $H$ the elements of $\bn_H$ are real. Since we have set $\hbar=1$, the elements of $\bn$ are frequencies. Working in terms of the normalized vector $\hat \bn_{H}=\bn_H/|\bn_H|$, which lives on the surface of the operator Bloch sphere, the time evolution under $H$ thence simply corresponds to a rotation about the axis $\bn_H$ at an angular velocity $|\bn_H|$. Let us define the two axes $\hat \bn_{0,1}:=\hat \bn_{H_{0,1}}$ and choose the coordinate system such that $\hat \bn_0=\hat \bz$, i.e. $\bB_0 \parallel \be_z$.

\section{Effective Hamiltonian}
\label{SEC:effective_H}

The conventional approach to obtain an approximate form of the effective Hamiltonian is via the Magnus expansion. In our case, there are however two disadvantages of going this route: first, the long and involved pulse sequence makes an analytical treatment of the non-ideal pulse case impossible and one would have to resort to a numerical treatment. Second, the period is very long compared to the typical time scales of the system and thus the convergence of the Magnus expansion towards the true effective Hamiltonian is not guaranteed.

We however still use the Magnus expansion to gain insight into the underlying principles of MAS, as well as to account for the emergence for the effective two-particle interactions in the case of idealized (instantaneous) pulses. For the Magnus expansion to converge, it firstly has to be performed within the toggling frame (Dirac picture w.r.t. all single-spin pulses). Secondly, we find that when using the full dipolar interaction (i.e. if the Magnus expansion is performed without using the secular approximation), the expansion does not converge to the correct solution at any Magnus order. This is shown in \reff{U_dist_Magnus_comparison}, where the operator distance $||\mU_{\mbox{\tiny exact}}-\mU_{\mbox{\tiny Magnus}}||$ of the exact unitary evolution and the unitary operator $\mU_{\mbox{\tiny Magnus}}=e^{-it \overline \mH_{\mbox{\tiny Magnus}}}$ obtained from the effective Magnus Hamiltonian to various orders is shown as a function of the inter-spin distance for two spins. However, this non-convergence is mitigated if the Magnus expansion is performed using the dipolar interaction within the secular approximation (discarding all terms leading to a flip of the total $s_{z,1}+s_{z,2}$ value in a large field $\bB_0=|\bB_0|\be_z$). In fact, in the latter case the effective Hamiltonian converges to the exact effective Hamiltonian computed with the full dipolar Hamiltonian (not using the secular approximation), as shown in \reff{U_dist_Magnus_comparison}.

\pic{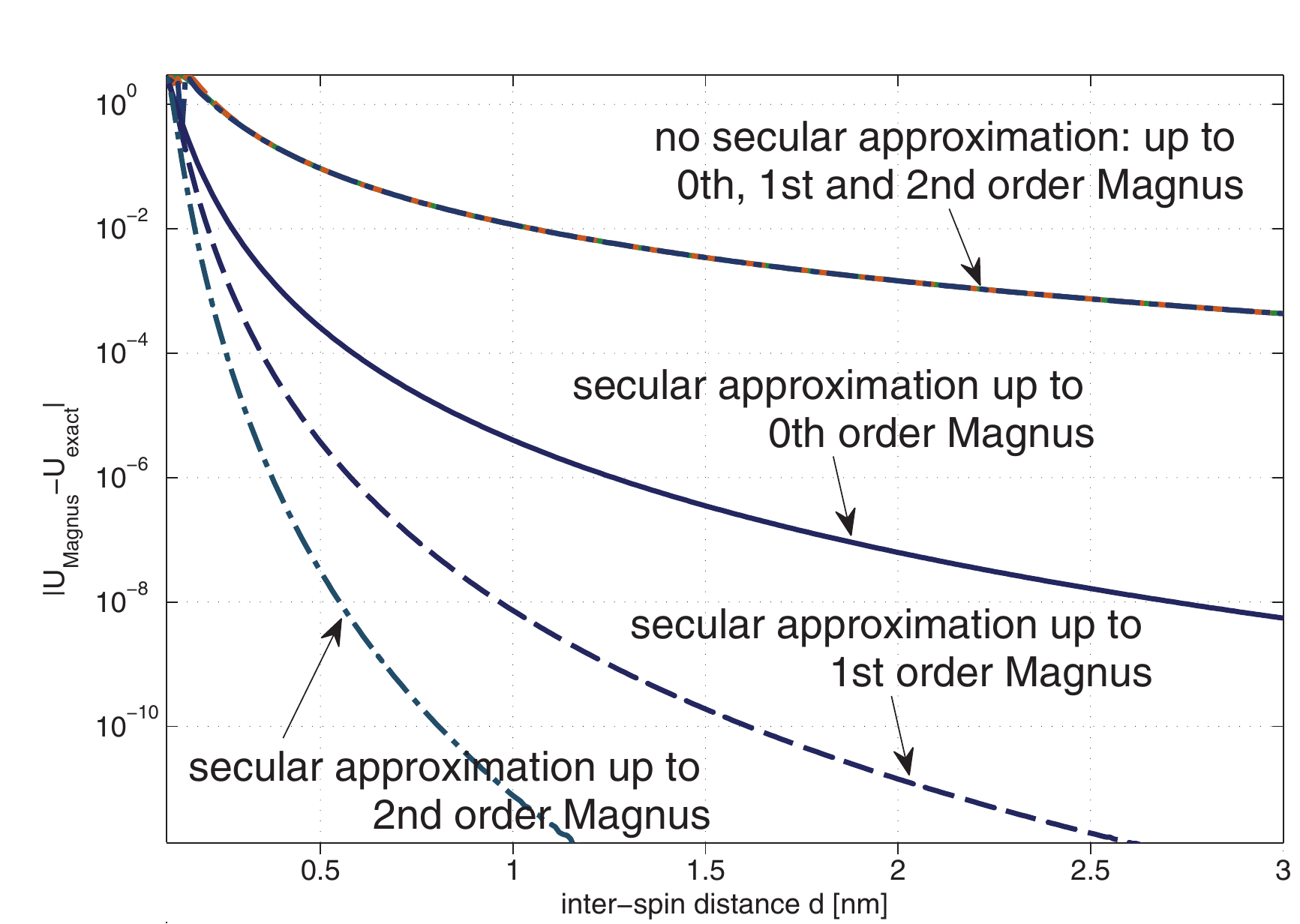}{\label{FIG:U_dist_Magnus_comparison} Distance between the exact and Magnus time evolution operators for two spins as a function of the inter-spin distance for $m=4$. Different orders of the Magnus expansion, both within and without the secular approximation are compared to the exact propagator. Surprisingly, the Magnus expansion only converges and has any validity within the secular approximation, also when compared to to the exact propagator without the secular approximation. }{\linewidth}

We therefore choose to calculate the effective Hamiltonian exactly (up to numerical errors). When performing a calculation of the propagator $\mU(T)$ (e.g. within an exact numerical simulation) in the Schr\"odinger picture, one however runs into the problem that simply calculating the effective Hamiltonian from $\overline \mH = i \ln(\mU(T)) / T$ gives rise to spurious interaction terms in some regimes, even if no physical interactions are present. These originate from the various branches of the logarithm within the eigenspace of each eigenvalue of $\mU(T)$, i.e. the non-uniqueness of the effective Hamiltonian associated with a given propagator, a problem that is never encountered within the Magnus expansion where the branch of every eigenvalues is followed continuously. In our case this problem is not easily mitigated, since the period $T$ is much longer than the inverse internal single-spin energy scales, implying that the \textit{correct} effective Hamiltonian (the one containing no effective interactions for non-interacting spins and obtained by the Magnus expansion if this converges) may lie far from the principal branch of the logarithm in various eigenstate sectors.

 To counteract these spurious contributions in the effective Hamiltonian originating from the single-particle dynamics, we work in the Dirac picture where $\mH_0$ is chosen to contain all single-spin contributions and thus only interaction effects contribute to the effective Hamiltonian $\overline \mH_I = i \ln(\mU_I(T)) / T$. Instead of finding the Dirac propagator $\mU_I$ by propagating the time-dependent equations of motion in the transformed rotating frame, this can also be determined from $\mU_I(T)= \mU_0^{-1}(T)\, \mU (T)$.

The effective Hamiltonian in the Dirac picture $\overline \mH_I = i \ln(\mU_I(T)) / T$ mitigates the problem of discontinuities arising from the multiple branches of the logarithm and it yields a meaningful effective Hamiltonian up to times $\propto 1/|\mH_{\mbox{\tiny int}}|$.

\section{Quantifying the Effective Interaction Strength}
\label{APP:strength_quantification}
Given an effective Hamiltonian $\overline \mH$ in numerically exact form and going beyond the Magnus expansion, we here shortly describe how we quantify the effective interaction strength between any pair of spins. For any single spin-$1/2$, the operators $\{\sigma_0:=\mathbbm 1, \, \sigma_x,\, \sigma_y,\, \sigma_z\}$ constitute a local operator basis, which is orthonormal with respect to the trace product $\tr(\sigma_k^\dag \sigma_l)/2=\delta_{k,l}$. An operator basis of the entire Hilbert space consisting of $M$ spins is given by all possible products of local spin basis operators $\mathcal B_{m_1,\ldots,m_M}=\sigma_{m_1}\otimes \ldots \otimes \sigma_{m_M}$. Any operator on the many-body space can easily be decomposed in this basis using the orthogonality
\spl{
\tr(\mathcal B_{m_1,\ldots,m_M}^\dag \mathcal B_{m_1',\ldots,m_M'})/2^M=\delta_{m_1,m_1'} \ldots \delta_{m_M,m_M'}.
}
Any two-spin operator describing an interaction between spin $j$ and $j'$ can then be written as a superposition of operators $B_{m_1,\ldots,m_M}$ where all $m_l=0$ except for $l=j$ or $l=j'$. As a measure for the interaction strength between spin $j$ and $j'$, we use 
\spl{
\overline {\mathcal S}_{j,j'}=\left[\sum_{m_j,n_j=1}^3 |\overline D_{m_j,n_{j}}^{j,j'}|^2 \right]^{1/2}
}
with the effective couplings being the coefficients of the respective basis operators above in the decomposition of the effective Hamiltonian
\spl{
\label{EQ:def_D}
\overline D_{m,n}^{j,j'}=\tr[(\sigma_{m}^j \otimes \sigma_{n}^{j'} ) \overline \mH]/2^M.
}
With this definition, $\overline {\mathcal S}_{j,j'}$ also corresponds to the trace norm of the interaction operator on the two-spin Hilbert space.

\subsection{Decoupling Ratio}
The analogous quantity ${\mathcal S}_{j,j'}$ can be defined for the bare dipolar two-spin interaction by simply replacing $\overline \mH$ with the bare dipolar Hamiltonian in \refe{def_D}. As defined in the main paper, the decoupling ratio $R_{j,j'}= \overline {\mathcal S}_{j,j'}/{\mathcal S}_{j,j'}$ is the relative strength of the effective interaction relative to the bare one. We point out that the maximum value that the decoupling ratio can take on in our system in the presence of a strong background field $\bB_0$ is 1/2. This originates from the fact that we perform our calculations using the full dipolar interaction in the bare Hamiltonian, i.e. not in the secular approximation. However, only the secular terms survive and contribute to the effective Hamiltonian. In quantifying the bare interaction strength ${\mathcal S}_{j,j'}$, it turns out that the secular components of the dipolar interaction Hamiltonian contribute exactly the same weight as the non-secular terms.

\section{Geometric Interpretation of the leading order Magnus Expansion}

\pic{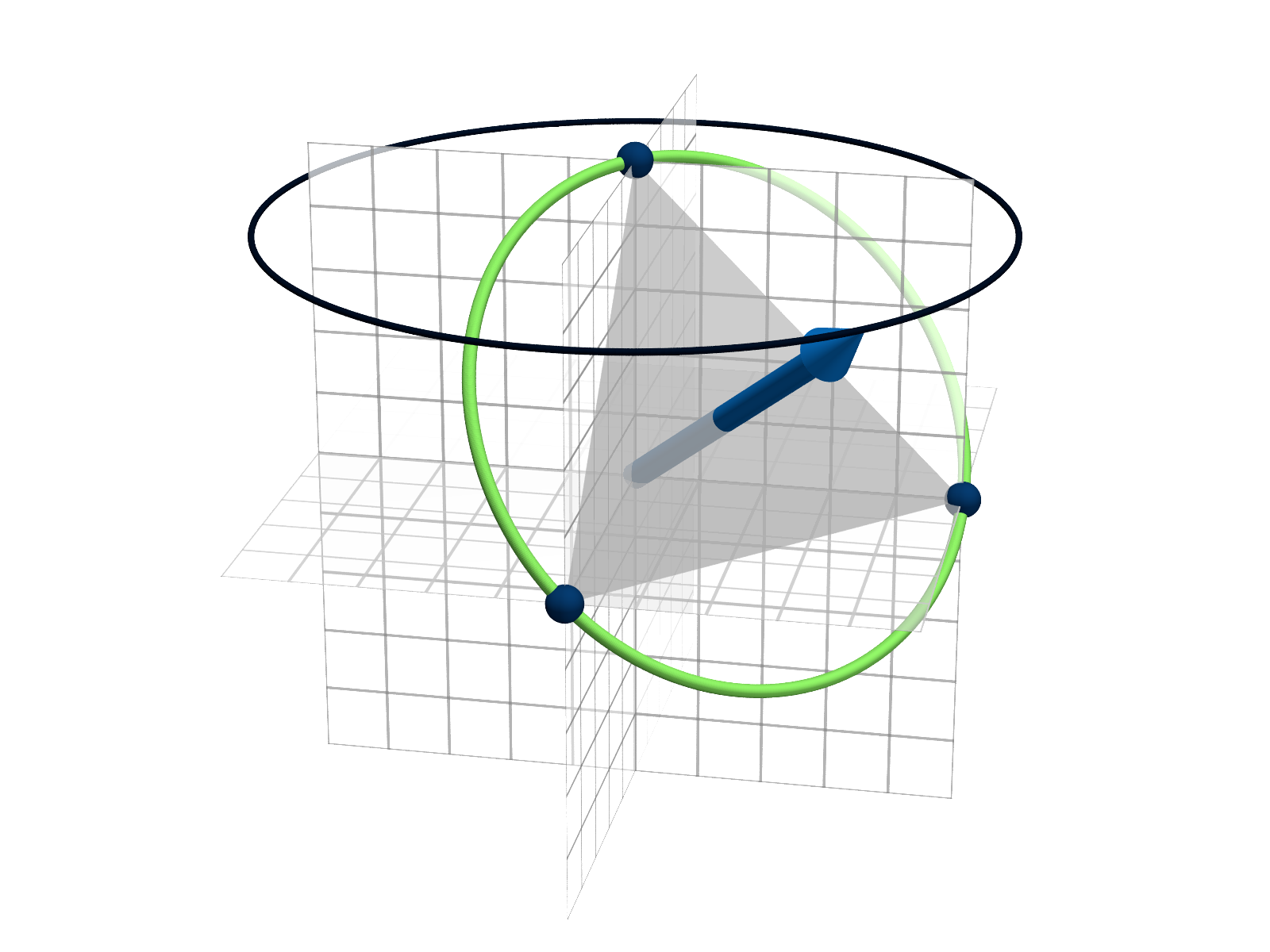}{\label{FIG:fig_magic_angle_discrete} Geometric representation of decoupling via magic angle spinning. A 3D vector $\bv$ in this space represents a 2-spin operator of the form $A=\sum_{d\in\{x,y,z\} } v_d I_{d}^{(1)} I_{d}^{(2)}$, such that the initial vectorial part of the dipolar interaction lies on the $z$-axis. If the rotation is performed around the magic axis (blue vector), this operator in the toggling frame is geometrically rotated around this very axis and takes on discrete points (blue spheres, shown here for $m=3$) on the green circle. The average Hamiltonian in the lowest oder Magnus approximation is given by the spatial average, indicated here by the intercept of the gray triangle with the rotation axis. Since the scalar part $\bI^{(1)} \cdot \bI^{(2)}$ of the dipolar Hamiltonian is also exactly represented by this geometric point, these two operators cancel, giving a geometric understanding of MAS decoupling. A closer inspection reveals that if the rotation axis (blue arrow) is itself rotated around the $z$-axis (lying anywhere on the black circle), only the rotated Hamiltonians are represented by geometric points reflected into the primary octant ($x,y,z \geq 0$). Hence, decoupling also occurs in this case, a property which also has to follow from the symmetry of the original physical system.}{\linewidth}

We will here show how the successive of $m$ instantaneous pulses corresponding to rotations by $2 \pi/m$ about the magic axis decouple the dipole Hamiltonian in the leading order of the Magnus expansion. First we note, that this procedure only works within the secular approximation of the dipole-dipole interaction Hamiltonian (i.e. if terms which change the total spin in $z$-direction are neglected), which requires a dominant magnetic field $B_0$. Within the secular approximation, the dipole Hamiltonian is
\spl{
\label{EQ:Hdd_appendix}
\Hdd \approx C [3 I_z^{(1)} I_z^{(2)} - \bI^{(1)} \cdot \bI^{(2)}],
}
where $C$ is a distance-dependent prefactor.

In the pulse sequence 
\begin{widetext}
\spl{
\label{EQ:total_pulse_seq_abs}
\Utot&=\left[ e^{-i \alpha H_\bn } e^{-i \tau \Hdd}  \right]^L\\
&= \underbrace{e^{-i \alpha H_\bn } e^{-i \tau \Hdd} e^{i \alpha H_\bn }}_{\mbox{$\Hdd$ rotated about $\bn$ by $\alpha$}} \; 
e^{-2i \alpha H_\bn } e^{i \tau \Hdd} e^{2i \alpha H_\bn } \ldots
\underbrace{e^{-iL \alpha H_\bn } e^{i \tau \Hdd} e^{iL \alpha H_\bn }  }_{\mbox{$\Hdd$ rotated about $\bn$ by $L\alpha$}} \: e^{-iL \alpha H_\bn } \\ 
 &= e^{-i \tau (e^{-i \alpha H_\bn }\Hdd e^{i \alpha H_\bn })} 
e^{-i \tau (e^{-2i \alpha H_\bn }\Hdd e^{2i \alpha H_\bn })} 
\ldots
e^{-i \tau (e^{-iL \alpha H_\bn }\Hdd e^{iL \alpha H_\bn })} \: e^{-iL \alpha H_\bn }, 
}
\end{widetext}
the individual pulses are assumed to be instantaneous compared to the wait time $\tau$ where the system evolves under $\Hdd$. The enclosing factors $e^{\pm i \alpha H_\bn }$ can simply be interpreted as rotation operators, rotating the secular $\Hdd$ away from the $z$ axis onto a sequential set of discrete points, shown as little spheres on a circle in \reff{fig_magic_angle_discrete}. These can be absorbed into the exponentials and the interpretation of the last line in Eq.~(\ref{EQ:total_pulse_seq_abs}) is an evolution of any state under a set of successively rotated secular dipole Hamiltonians on a cone. Note that this is simply a rewriting and is exact. To understand the decoupling to leading order, we consider the effective Hamiltonian via the Magnus expansion, setting $\alpha=2 \pi /L$ to generate a periodic Hamiltonian and looking at one period $\alpha L$.

Let us first consider a term of the form $e^{-ik \alpha H_\bn }\Hdd e^{ik \alpha H_\bn }$: the application of the two unitary operators simply correspond to a rotation of $\Hdd$ around the axis $\bn$ by an angle $k \alpha$. Looking at the two terms of $\Hdd$ in Eq.~(\ref{EQ:Hdd_appendix}), we notice that the second term is simply a scalar and remains invariant under any rotation $e^{-ik \alpha H_\bn } \bI^{(1)} \cdot \bI^{(2)} e^{ik \alpha H_\bn } = \bI^{(1)} \cdot \bI^{(2)}$. The first term transforms as a dyadic product of two vector operators. The transformation of any vectorial operator $A:=m_x I_x + my I_y + m_z I_z$ can be conveniently expressed in the following way: if $\bmm$ parametrizes the initial operator, the rotated operator is parametrized by $\bmm'=\mR \bmm$, where $\mR$ is the three-dimensional representation of the respective rotation, i.e.
\spl{
e^{-i \alpha H_\bn }\Hdd e^{i \alpha H_\bn } = \bmm'^\dag \cdot \bI = \bmm^\dag \, \mR^\dag \,  \bI.
}

The effective Hamiltonian $\overline H$ (defined via $\Utot(T)= e^{-iT \overline H}$) to lowest order in the Magnus expansion is thus
\al{
\label{EQ:H0_Magnus}
\overline H_0 &=  \vvec{I_x^{(1)} \\ I_y^{(1)} \\ I_z^{(1)}}^t \mM(\bn) \vvec{I_x^{(2)} \\ I_y^{(2)} \\ I_z^{(2)}}\\
\mM(\bn)&=\frac 1 m \sum_{k=1}^m \mathcal{R}^m(k \alpha, \bn)  \vvec{0& 0& 0 \\0& 0& 0\\0& 0& 1}   {\mathcal{R}^\dag}^m(k \alpha, \bn) 
}
It is useful to analyze each of the terms in the eigenbasis of $\mathcal{R}(k \alpha, \bn)$. Note that the $\mathcal{R}'s$ are the three-dimensional representation of the rotation group. Although all elements can be chosen real, they are generally only diagonalizable over $\mathbb{C}$. The rotation axis $\bn$ is always an eigenvector to eigenvalue $1$. In the orthogonal subspace a representation of $\mathcal{R}(k \alpha, \bn)$ is of the form of a 2D rotation $\vvec{\cos(k \alpha)& -\sin(k \alpha)\\ \sin(k \alpha) & \sin(k \alpha)}$ with the eigenvectors $(1, \, -i)^t$ and $(1, \, i)^t$ to the eigenvalues $e^{i k \alpha}$ and $e^{-i k \alpha}$ respectively. 
Let $\bv_\pm$ denote the 3d counterparts of these eigenvectors, which, together with $\bv:=\bn$ form an orthonormal basis. Since $\mR$ is unitary orthogonal, the eigenvectors to the mutually conjugated eigenvalues are related by conjugation $(v_+)^*=v_-$.

Inserting the spectral decomposition of $\mR$ and since only the eigenvalues of $\mR$ and not the eigenvectors depend on the angle $k\alpha$, the coupling matrix can be expressed as 
\begin{widetext}
\spl{
\mM( \bn)&=\frac 1 m \sum_{k=1}^m [\bv \bv^\dag + e^{i k \alpha} \bv_+ \bv_+^\dag+ e^{-i k \alpha} \bv_- \bv_-^\dag] \be_z \be_z^\dag [\bv \bv^\dag + e^{-i k \alpha} \bv_+ \bv_+^\dag+ e^{i k \alpha} \bv_- \bv_-^\dag]\\
&= |\bv \cdot \be_z|^2 \bv \bv^\dag + |\bv_+ \cdot \be_z|^2 \bv_+ \bv_+^\dag+ |\bv_- \cdot \be_z|^2 \bv_- \bv_-^\dag,
}
\end{widetext}
where we have used that all oscillating terms vanish over a full period $\sum_{k=1}^L e^{i j k \alpha}=L \, \delta_{j,0}$. Parametrizing the rotation axis in spherical coordinates $\bn=(\sin\theta \cos \varphi, \, \sin\theta \sin \varphi, \, \cos\theta)$, together with the completeness relation 
\spl{
\bv \bv^\dag+ \bv_+ \bv_+^\dag+\bv_- \bv_-^\dag=\mathbbm{1}_3
}
and the property $|\bv_+^\dag \cdot \be_z|=|\bv_-^\dag \cdot \be_z|$, we have
\spl{
|\bv_+^\dag \cdot \be_z|^2=|\bv_+^\dag \cdot \be_z|^2=\frac{\sin^2 \theta}{2}.
}
If $\bn$ is tilted at the magic angle (with respect to $\be_z$), i.e. $\theta=\thetaM=\arctan \sqrt 2$, we have $\cos^2 \thetaM=1/3$ and $\sin^2 \thetaM=2/3$, leading to $\mM( \bn_M)={\mathbbm{1}_3}/{3}$. The averaged non-scalar term thus exactly cancels the scalar part in Eq.~(\ref{EQ:H0_Magnus}), thus decoupling the two spins to this order. We note that this decoupling is independent of the relative orientation of the two dipoles. The decoupling can furthermore be extended to the next order by traversing the cone in successively opposite directions.

\section{Decoupling Performance under MAS}
\label{APP:decoupling_peaks}

\pic{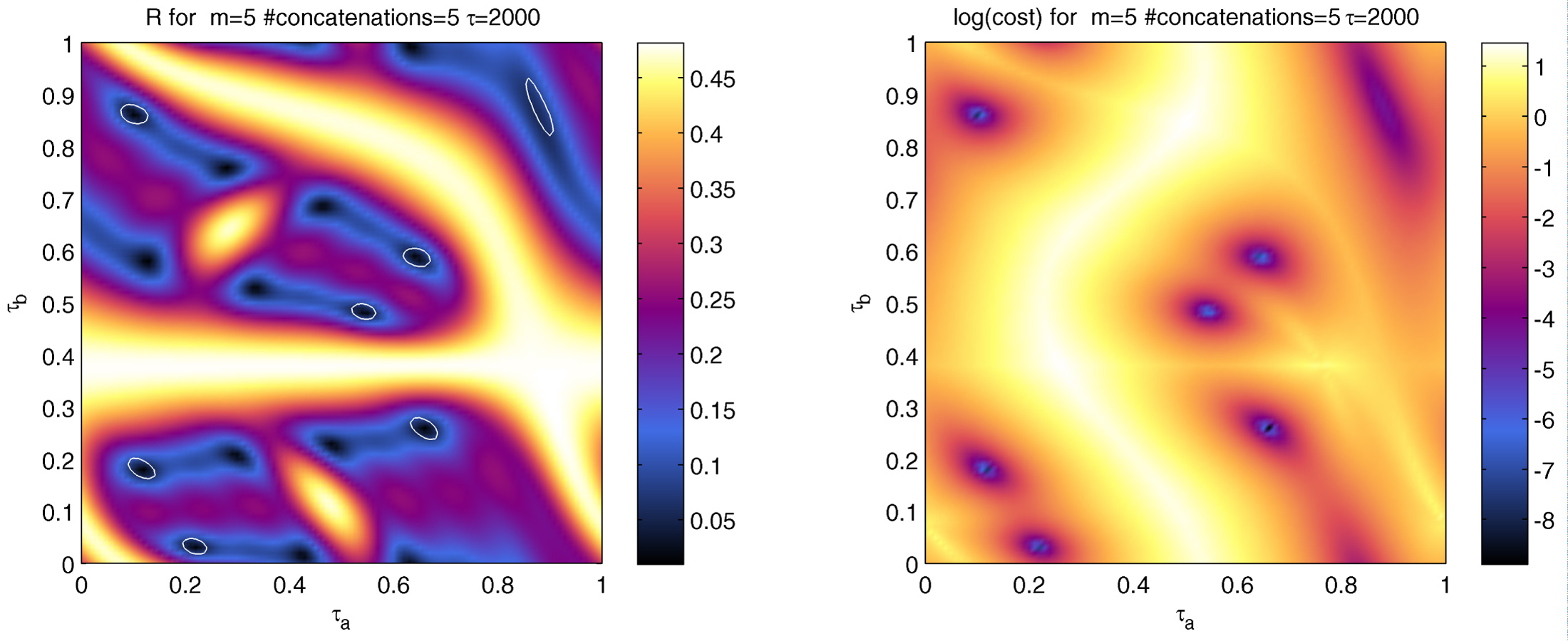}{\label{FIG:decoupling_peaks} That our concatenation procedure and effective MAS is the actual cause of the observed decoupling is demonstrated by the minima of a cost function (shown in b) always leading to minima in the decoupling ratio (shown in (a), with the minima from (b) shown as white circles). The cost function in (b) measures the deviation of the rotation in the concatenated propagator from the ideal magic angle rotation operator in the single particle sector. }{\linewidth}

We generally find, as reflected in \reff{decoupling_peaks}(a), that every optimal magic angle rotation unitary leads to decoupling, but not vice versa; i.e. the effective magic angle spinning protocol is a sufficient criterion for decoupling, but not necessary. This result is remarkable and not clear a priori, since the construction time for the effective rotation is not small on the inverse interaction energy scale and the usual derivation of MAS relying on the lowest order Magnus expansion does hence not apply.
It should be mentioned that the structure of $\mathcal R(\tau_a, \tau_b)$ strongly depends on the parameters $\tau_1$, $m$, $N_c$ and \reff{decoupling_peaks}(a) is just one example.

\section{Exploiting the wait time $\tau$ between rotation pulses} \label{SEC:wait_time}
Before we have neglected the additional degree of freedom of free evolution under $H_0$ between the individual rotation pulses. Here, we show how the wait time $\tau$ can be used as an additional, independent \emph{knob} to adjust and potentially increase the decoupling ratio $\mR$ by many orders of magnitude, as shown in \reff{wait_time_combined}.

\picwide{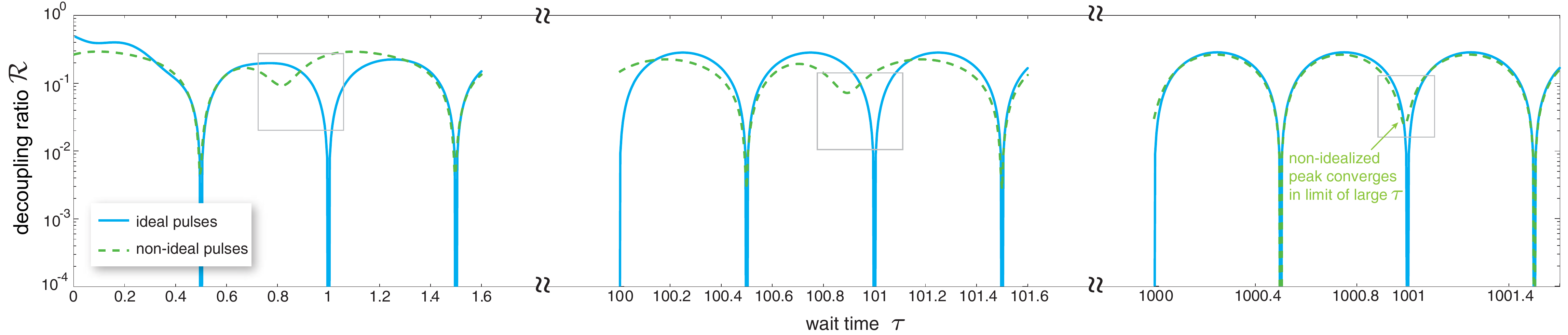}{\label{FIG:wait_time_combined} Comparison of the dependence of the decoupling ratio (note the log-scale) on the wait time $\tau$ for $m=3$ for ideal pulses (blue solid line) vs non-ideal pulses constructed using our procedure for $m=3$. For ideal pulses there is perfect decoupling ($\mR=0$) at both integer and half-integer multiples of the Larmor time. For non-ideal  pulses, the peak structure is not very pronounced, especially for the decoupling peak at integer times. In the limit of large $\tau$, however, where the construction time of the non-ideal pulse (a few hundred Larmor times here) becomes small compared to $\tau$, the non-ideal decoupling ratio structure converges to the ideal limit. In all regimes, the half-integer peak is much more pronounced than the integer peak for the non-ideal pulse sequence.}{\linewidth}

We here focus on the case $m=3$. In addition to the expected decoupling peak at integer multiples of the Larmor time which arises as the spins simply precesses an integer number of times around $\bB_0$ during $\tau$, recovering the initial state (up to a possible global sign).

The peak at half-integer rotation times under $B_0$ can be understood within the secular approximation. We go into the interaction picture with respect to all single spin contributions, i.e. both $B_0$ and the time-dependent pulses, such that only the dipole interaction in this rotating frame generates the dynamics. The non-interacting propagator can be written segment-wise for $(m-1)\tau < t < m\tau$, where $\tau$ is the wait time between pulses and $m\in \mathbb N$ as $U_0(t)=[e^{i \mH_0 \tau} U_\bn^\dag(\varphi)]^m$ and the interaction operator in the rotating frame becomes
\spl{
\Hint(t)=&[e^{i \mH_0 \tau} U_\bn^\dag(\phi)]^m  \, e^{i \mH_0 (t-m\tau)}  \Hint  e^{-i \mH_0 (t-m\tau)}\\ & \times [ U_\bn(\phi) e^{-i \mH_0 \tau}]^m.
}
In the secular approximation, where $[\Hint^{(\mbox{\tiny sec})}, H_0]=0$ the central part reduces to the time-independent interaction $e^{i \mH_0 (t-m\tau)}  \Hint^{(\mbox{\tiny sec})}  e^{-i \mH_0 (t-m\tau)}=\Hint^{(\mbox{\tiny sec})}$ and
\spl{
\Hint^{(\mbox{\tiny sec})}(t)=[e^{i \mH_0 \tau} U_\bn^\dag(\phi)]^m  \, \Hint^{(\mbox{\tiny sec})} [ U_\bn(\phi) e^{-i \mH_0 \tau}]^m.
}
Thus, the secular dipole interaction is piecewise constant in the rotating frame with only discrete, instant rotations $[ U_\bn(\phi) e^{-i \mH_0 \tau}]^m$ acting at times $m\tau$.
The perfect decoupling at integer rotation times $\tau=2\pi l/|B_0|$, with $l \in \mathbb N$ is immediately clear, as $e^{-i \mH_0 \tau}=\mathbbm 1$ drops out of the problem at these times in the rotating frame.

In essence, the second non-integer decoupling peak can be explained by the composite rotation $U_\bn(\phi) e^{-i \mH_0 \tau}$ also being a magic axis, possibly with a different azimuth angle. Specifically, we start with the sequence $m=3$, where the non-integer decoupling peak appears at $\tau=2\pi(l+\frac 1 2)/|B_0|$ and has a simple geometric interpretation. Instead of analyzing the two-dimensional spin-1/2 representation of the rotation, we describe the procedure in the three-dimensional, locally isomorphic representation. Any 3D rotation operator other than the unit operator, possesses a single real eigenvalue and the associated eigenvector is the rotation axis. Thus, to prove that a given vector $\ba$ is the effective overall rotation axis of a set of concatenated rotations, it is sufficient to prove that starting from $\ba$, it is mapped back onto itself after applying the sequence of rotations.

As before, the fractional rotation operator $U_\bn(\phi)$ rotates around the magic axis $\bn=[\sqrt 2, \, 0, \, 1]/\sqrt{3}$ by a flip angle $\phi=2\pi / 3$. The rotation axis of $U_{\bn'}(\phi')=U_\bn(\phi) e^{-i \mH_0 \tau}$ is also titled at the magic angle relative to $B_0 \parallel \be_z$, but in the $y$-$z$-plane $\bn'=[0,\, \sqrt 2, \, 1]/\sqrt{3}$, as we will now prove. The eigenvectors of a rotation operator correspond to its rotation axis. Thus, if we start with $\bn'=[0,\, \sqrt 2, \, 1]/\sqrt{3}$, the first part of the rotation by $\pi$ around $\be_z$ will map this onto $e^{-i \mH_0 \tau}\bn'=[0,\, -\sqrt 2, \, 1]/\sqrt{3}$. The rotation around $\bn$ will subsequently keep the vector in the plane $A=\{\bv \; | \; (\bv - \bn/\sqrt 3)\cdot \bn =0   \}$ perpendicular to $\bn$, containing $\bn'$. It can now easily be seen that the subsequent rotation $U_\bn(\phi) $ on $\bw=[0,\, -\sqrt 2, \, 1]/\sqrt{3}$ by $\phi=2\pi / 3$ rotates $\bw$ back onto $\bn'$, as the angle subtended by $(\bw-\bn/\sqrt 3)$ and $(\bn'-\bn/\sqrt 3)$ (within the plane $A$) is 
\spl{
\phi'&=\arccos\left[ \frac{(\bw-\bn/\sqrt 3)\cdot (\bn'-\bn/\sqrt 3)}{|\bw-\bn/\sqrt 3| \, |\bn'-\bn/\sqrt 3|}  \right]\\
&=\arccos\left(-\frac 1 2 \right)=\frac{2\pi}{3}.
}
Therefore decoupling can also be achieved at half-integer Larmor rotation wait times between pulses, as the resulting effective rotations are also at the magic angle, albeit around an axis rotated $90^\circ$ around $\be_z$ relative to the axis of the bare rotation pulses. This does not hold for intermediate wait times $\tau$ (neither integer nor non-integer Larmor times), as the effective rotation axis is not titled at the magic angle.

\section{Enhanced Effective Two-Spin Interactions}
\label{SEC:tertiary_spin_induced_int}
As shown in the exact numerical simulations, tertiary spins can significantly enhance the effective two-spin interactions between a given pair by one to two orders of magnitude. Here we show how this phenomenon can also be understood from the Magnus expansion, with the second order terms being the lowest non-vanishing contributor. As before by construction, the zeroth order terms in the Magnus expansion in the toggling frame average to zero, while the first order terms involving commutators of the form $[ \overline \mH_{j_1,j_2}(t1), \overline \mH_{j_2,j_3}(t1) ]$ give rise to effective three spin interactions. The third order terms, containing nested commutators $[ \overline \mH_{j_1,j_2}(t1), [ \overline \mH_{j_2,j_3}(t1), \overline \mH_{j_1,j_3}(t1) ] ]$ in turn again contribute to the effective two-spin interaction between spins $j_1$ and $j_2$, i.e. an interaction mediated by tertiary spin $j_3$.

\pic{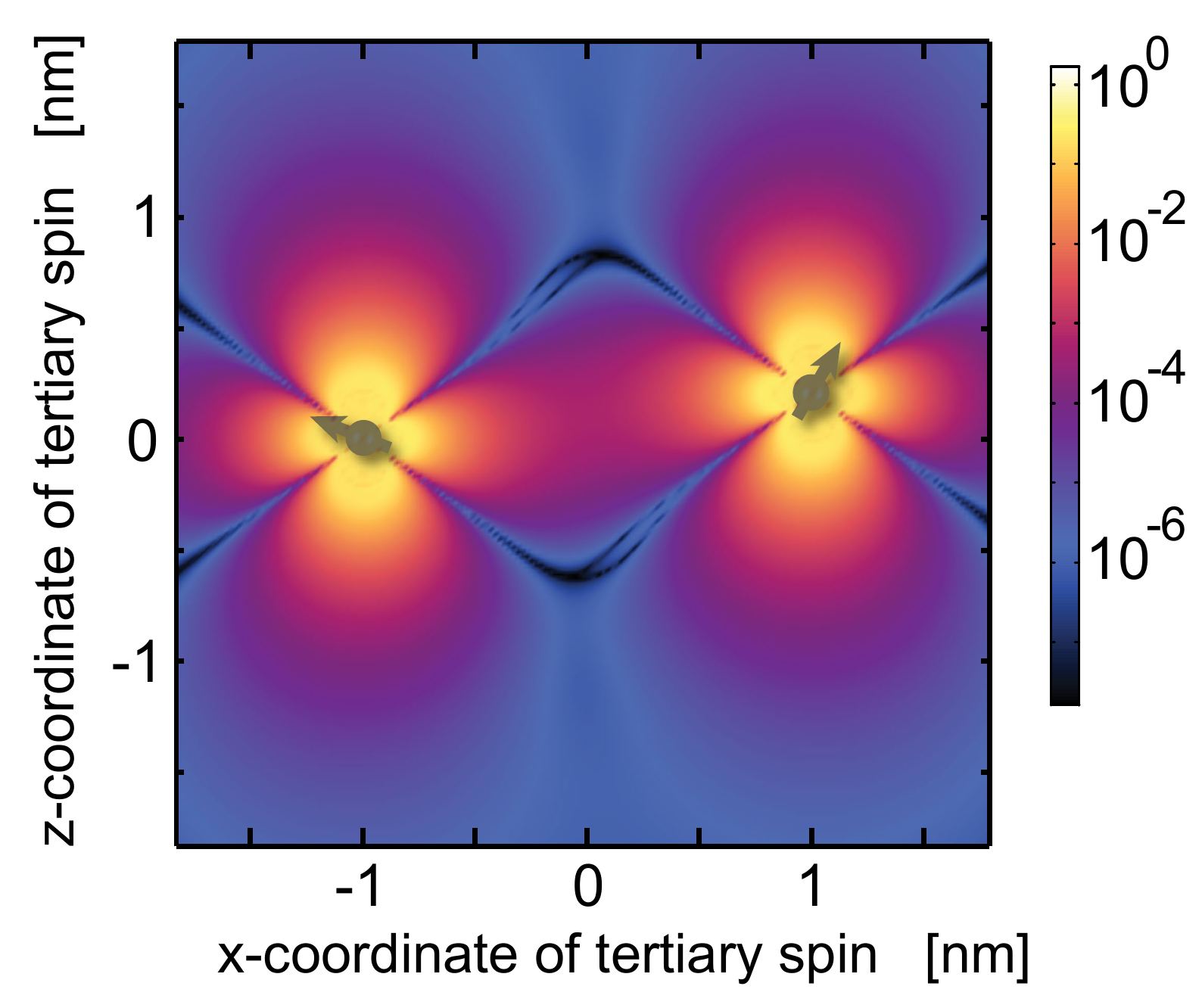}{\label{FIG:induced_interaction_2d_plane} The decoupling ratio (note the logarithmic color scale) of two spins fixed at positions $(-1, 0, 0)^t$nm and $(1, 0, 0.2)^t$nm as a function of the position of a tertiary spin, which affects the effective two-spin interaction. The decoupling ratio is shown here for idealized pulses with $\tau=1000$ and $m=3$ in the plane oriented in the $e_z$ direction and containing the two spins (their position is chosen arbitrarily). }{0.9\linewidth}

The magnitude of the total effective two-spin interaction is shown as a function of the tertiary spin's position in the $y=0$ plane in \reff{induced_interaction_2d_plane}. Relative to the MAS-suppressed case, the effective two-spin interaction can be increased by more than three orders of magnitude by the tertiary spin. A closer inspection reveals that the magnitude does not exceed the bare two-spin interaction though.

\section{Dependence of the Effective Two-Spin Interaction on Tertiary Spins}

When extracting the two-spin part of the interactions $\overline \mH_{j,j'}$ from the full effective Hamiltonian $\overline \mH$, the former may depend on tertiary spins other than $j$ and $j'$. To elucidate this dependence, we consider an equispaced 1D spin chain consisting of an even number $L$ of spins. Here we focus on $\overline \mH_{j,j'}(L)$ of the two central spins $j$ and $j+1$, which is now a function of the interspin distance $|\br_j-\br_{j+1}|$ and $L$. One would expect $\overline \mH_{j,j'}(L)$ to converge asymptotically with increasing $L$. To demonstrate and quantify this convergence, we plot $\| \mH_{j,j'}(L) - \mH_{j,j'}(L_{\mbox{\tiny max}})  \|$ in \reff{nn_H_dep_on_no_neighbors_no_NV} for $L_{\mbox{\tiny max}}=8$, where $\| \ldots \|$ is the Frobenius norm.

\pic{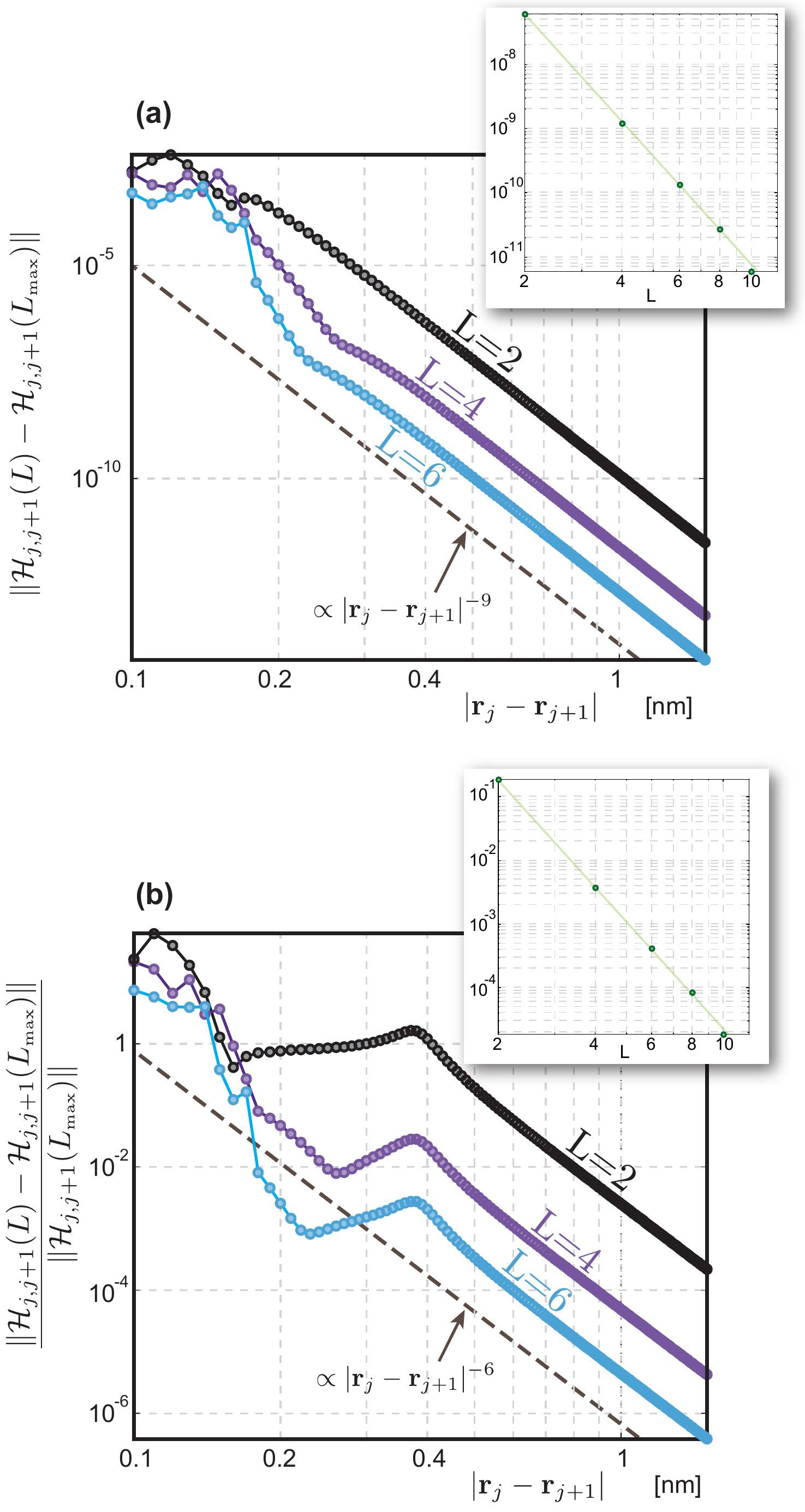}{\label{FIG:nn_H_dep_on_no_neighbors_no_NV} Dependence of effective two-spin interaction on tertiary spins. We consider for simplicity no NV center control, just global pulses, and $\tau=100.5$, $m=3$, $N_c=5$. The two insets show the dependence of the respective quantity on $L$ at $|\br_j-\br_{j+1}|=0.5$nm and $L_{\mbox{\tiny max}}=12$, which scales approximately as $\propto L^{-6}$.}{0.8\linewidth}

For the nuclear spin coupling strength of $^{13}$C, we find that for spacings above a critical distance $d_{\mbox{\tiny crit}} \approx 0.15\mbox{nm}$ (at this distance $||\mH_{j,j'}||/|| \mH_0||\approx 0.002$, where both Hamiltonians are taken from the two-body space), which is coincidentally very close to the carbon-carbon bond length, $\overline \mH_{j,j'}(L)$ converges with $L$. Equivalently with increasing inter-spin distance, not only the magnitude of $\mH_{j,j'}(L)$ decreases, but also the difference $\| \mH_{j,j'}(L) - \mH_{j,j'}(L_{\mbox{\tiny max}})  \|$ scales as $\propto |\br_j-\br_{j+1}|^{-9}$.

The relative accuracy $\| \mH_{j,j'}(L) - \mH_{j,j'}(L_{\mbox{\tiny max}})  \| / \|  \mH_{j,j'}(L_{\mbox{\tiny max}})  \|$ also decreases as $\propto |\br_j-\br_{j+1}|^{-6}$ as shown in \reff{nn_H_dep_on_no_neighbors_no_NV} (b), indicating that the effective two-spin interaction can be calculated within a small cluster with increasing accuracy and tertiary spins become less relevant as the inter-spin distance increases.

\section{Protocol Parameters for Fig.~(4)}
\label{SEC:parameters_hexagonal_plots}

We here give the numerically optimized parameters used in Fig.~(4) of the main paper. All times are given in units of $2\pi/|\gamma_I \bB_0|$.
\begin{center}
    \begin{tabular}{ | c | c | c | c | c | c |}
    \hline
    NV depth& $\quad m \quad$ & $\tau$ & $\tau_a$ & $\tau_b$ & $\tau_1$ \\ \hline
    10nm & 24 & $\;$ 500.5 $\;$&$\;$ 0.076592 $\;$&$\;$ 0.49398 $\;$& $\;$0.42859$\;$ \\ \hline    
    6nm & 24 & 100.5 & 0.035455 & 0.34683 & 0.46816 \\ \hline    
    10nm & 40 & 500.5 & 0.12941 & 0.19712 & 0.49611 \\ \hline    
    5.7nm & 22 & 350 & 0.1128 & 0.75463 & 0.30256 \\ \hline    
    \end{tabular}
\end{center}

\end{appendix}

\end{document}